\documentclass[12pt]{article}

\usepackage{graphicx}
\usepackage{epstopdf}
\usepackage{pstool}
\usepackage{jheppub} 

\usepackage{float}    %
\usepackage{dcolumn}   
\usepackage{bm}        
\usepackage{amssymb}   

\usepackage{float}
\usepackage{subcaption}
\usepackage{color}
\usepackage[dvipsnames]{xcolor} %

\usepackage{slashed}
\usepackage[utf8]{inputenc}
\usepackage{multirow}
\usepackage{footnote}
\usepackage{amsmath}
\usepackage{fancyhdr}
\allowdisplaybreaks[4]
\usepackage{accents}
\graphicspath{{Figs/}}
\newlength{\dhatheight}

\def\figureautorefname~#1\null{Fig.\,#1\null}
\def\tableautorefname~#1\null{Tab.\,#1\null}

\def\equationautorefname~#1\null{Eq.\,(#1)\null}

\title{ Dark Matter and Electroweak Phase Transition in the $Z_2$ Symmetric Georgi-Machacek Model}
\author[a,b]{ Chih-Ting Lu,}
\author[a,b]{ Yongcheng Wu,}
\author[a]{ Siyu Xu}
\affiliation[a]{Department of Physics and Institute of Theoretical Physics, Nanjing Normal University, Nanjing, 210023, China}
\affiliation[b]{Nanjing Key Laboratory of Particle Physics and Astrophysics, Nanjing, 210023, China}
\emailAdd{ctlu@njnu.edu.cn}
\emailAdd{ycwu@njnu.edu.cn}
\emailAdd{siyuxu@njnu.edu.cn}
\preprint{$\begin{gathered}\includegraphics[width=0.05\textwidth]{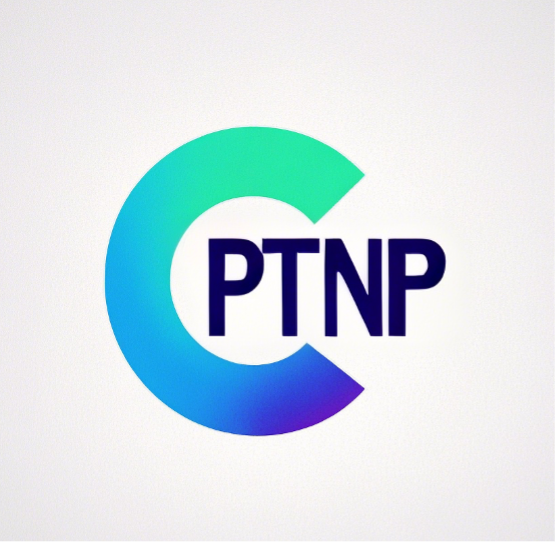}\end{gathered}$\,CPTNP-2025-003}

\abstract{
We present a comprehensive investigation of the $Z_2$ symmetric Georgi-Machacek (GM) model, focusing on the dark matter (DM) in the model and the electroweak phase transition (EWPT). Our analysis encompasses multiple detections for the DM candidates, including collider searches at the LHC and LEP, the direct detection and indirect detection. Furthermore, we also explore the possibility of a first-order EWPT in this framework. The gravitational wave (GW) generated from the first-order EWPT also provides a detection method for the parameter space in the $Z_2$ symmetric GM model providing viable DM candidate. It is found that the current DM searches, especially the direct detection, provide strong constraints on the parameter space, while the GW signal can be complementary around the Higgs resonant region.
}

\begin{document}
\titlepage
\maketitle

\newpage

\flushbottom

\section{Introduction}

The quest to understand the fundamental nature of dark matter (DM) and the physics behind electroweak symmetry breaking (EWSB) are among the most important targets in particle physics. The Standard Model (SM) extended with extra scalars are important benchmarks to be explored in such directions. The extra scalars, under certain circumstance, can be treated as DM candidate, while they also play crucial role in the EWSB, altering the electroweak phase transition (EWPT) patterns.

The Georgi-Machacek (GM) model~\cite{Georgi:1985nv,Chanowitz:1985ug,PhysRevD.42.1673} is one of such extensions which was proposed with the goal of exploring whether EWSB patterns different from those predicted by the SM still remain viable. It extends the SM by introducing two additional scalar triplets while preserving the custodial symmetry which maintains $\rho=1$ in accordance with the experimental observations. It can also provide enhanced couplings between the Higgs and the gauge bosons compared to the one in the SM. This is achievable only in the extensions with the scalar representation of ${\rm SU}(2)_L$ larger than the doublet. The fermiophobic fiveplet containing doubly charged component is also of great interests as benchmark for the searches of new physics at the LHC~\cite{CMS:2012dun,ATLAS:2015edr,CMS:2017fhs,ATLAS:2017uhp,CMS:2019qfk,ATLAS:2021jol,CMS:2021wlt,ATLAS:2022zuc,ATLAS:2024txt}. Recent studies about the general GM model can be found in~\cite{Shang:2025cig,Ashanujjaman:2025puu,Du:2025eop,Ghosh:2025ebc,Ghosh:2025htt,Lu:2024ade,Xu:2024yuy,Bhattacharya:2024gyh,Chen:2023bqr,Chakraborti:2023mya,Song:2022jns,Bairi:2022adc,Wang:2022okq,Ismail:2020kqz,Ismail:2020zoz,Ghosh:2019qie,Yang:2018pzt,Logan:2018wtm,Chiang:2018cgb,Sun:2017mue,Li:2017daq,Hartling:2014zca}.

The GM model was originally studied with $Z_2$ symmetry until the trilinear terms have been introduced in Ref.~\cite{Aoki:2007ah}, where the authors also found that in the $Z_2$ symmetric GM model, when the triplets obtain the vacuum expectation values (VEVs),
there is no decoupling limit in the model, such that the current LHC searches might already exclude the entire parameter space~\cite{Aoki:2007ah} similar to that in $Z_2$ symmetric two-Higgs-doublet model (2HDM)~\cite{Chowdhury:2017aav,Cacchio:2016qyh}. However, this conclusion only holds when the triplets obtain VEVs. For the case where the triplets have no VEVs, the $Z_2$ symmetry is preserved by the vacuum, the decoupling limit can be naturally achieved. In such case, the scalars from the triplets can be DM candidates. The DM in extensions of GM model have been studied previously~\cite{Campbell:2016zbp,Pilkington:2017qam,Chen:2020ark}, while recent studies on $Z_2$ symmetric GM model can be found in Refs.~\cite{Azevedo:2020mjg,deLima:2022yvn}, we will focus on the DM possibility within $Z_2$ symmetric GM model.

On the other hand, the extra scalars introduced in the GM can modify the EWPT patterns through their interactions with the SM Higgs. The EWPT in GM model has been studied in Refs.~\cite{Zhou:2018zli,Chen:2022zsh,Chiang:2014hia,Bian:2019bsn} without $Z_2$ symmetry, where the trilinear couplings among scalars in the potential play an important role to achieve the first-order EWPT. However, in the $Z_2$ symmetric GM model, the trilinear couplings are eliminated due to the $Z_2$ symmetry. The EWPT pattern might be changed accordingly. Hence, it would also be important to also investigate the EWPT possibility in the $Z_2$ symmetric GM model, especially when the $Z_2$ symmetry is also related to the DM candidate in GM model.

The remaining sections of the paper are organized as follows. In Sec.~\ref{sec:Z2GM}, we briefly revisit the GM model with the $Z_2$ symmetry, discussing the tree-level mass spectrum, theoretical constraints, and the conditions of the EWPT on the model parameters. In Sec.~\ref{sec:relic_density}, we discuss the relic density of the DM candidates followed by the discussion of the current experimental constraints from collider experiments, direct detection, and indirect detection experiments and the detection of gravitational wave signals originating from the first-order EWPT in Sec.~\ref{sec:constraints}. Finally, we conclude in Sec.~\ref{sec:conclusion}.

\section{Georgi-Machacek model with $Z_2$ symmetry}
\label{sec:Z2GM}

The GM model~\cite{Georgi:1985nv,Chanowitz:1985ug,PhysRevD.42.1673} extends the SM with two extra electroweak triplet, one real and one complex. With the specific arrangement of the triplets, the GM model preserve the custodial symmetry at the tree-level even if the triplet acquires large VEVs. However, it has been argued that the $Z_2$ symmetric GM model will not have decoupling limit if the $Z_2$ symmetry is spontaneously broken~\cite{Aoki:2007ah}. The masses of the scalars from the electroweak triplet do have upper limit which has already been covered by current LHC experiments. However, such situation will not happen if the $Z_2$ symmetry is preserved. In such case, the lightest scalar from the electroweak triplet, which is odd under the $Z_2$ symmetry, will be stable and can be a good DM candidate. The DM properties are highly correlated with the properties of the scalar potential. On the other hand, even if the $Z_2$ symmetry is preserved at zero temperature, it can be broken at high temperature. The transition from the other phase to $Z_2$ symmetric phase can be first-order, which can generate detectable gravitational wave signals. The property of the transition is also determined by the shape of the scalar potential and hence can be related with the DM properties. In this section, we will briefly discuss the parameterization of the $Z_2$ symmetric GM model, the theoretical constraints, the mass spectrum of the scalars and further the Electroweak phase transition in the model.

\subsection{Parameterization of the scalar potential in dark matter phase vacuum}

The scalar sector of the GM model contains one doublet $(\phi^+,\phi^0)^T$ and two triplets, one real $(\xi^+,\xi^0,\xi^-)^T$ and one complex $(\chi^{++}, \chi^+,\chi^0)^T$. The doublet and triplets can be arranged into a bi-doublet $\Phi$ and a bi-triplet $X$ to explicitly show the preservation of the custodial ${\rm SU}(2)_C$ symmetry, which are
\begin{align}
\Phi = \begin{pmatrix}
    \phi^{0 *} & \phi^{+} \\
    -\phi^{+*} & \phi^{0}
\end{pmatrix}, \quad X = \begin{pmatrix}
    \chi^{0 *} & \xi^{+} & \chi^{++} \\
    -\chi^{+*} & \xi^{0} & \chi^{+} \\
    \chi^{++*} & -\xi^{+*} & \chi^{0}
\end{pmatrix}.
\end{align}
In the $Z_2$ symmetric case, only the doublet $\Phi$ obtains the vacuum expectation value (VEV), which leads to the following form of the vacuum configuration
\begin{align}
    \langle \Phi \rangle = \frac{1}{\sqrt{2}} \begin{pmatrix}
        v_{\phi} & 0 \\
        0 & v_{\phi}
    \end{pmatrix}, \quad \langle X \rangle = \begin{pmatrix}
        v_{\chi} & 0 & 0 \\
        0 & v_{\chi} & 0 \\
        0 & 0 & v_{\chi}
    \end{pmatrix} = 0_{3\times 3}
\end{align}
Here, $v_\phi \equiv v \approx 246$ GeV is the total VEV at zero temperature. The fact that the triplets do not acquire a VEV ensures the existence of a viable DM candidate. This configuration is referred to as the "dark matter phase". The most general scalar potential in the $Z_2$ symmetric GM model can be written as~\cite{Hartling:2014zca,Azevedo:2020mjg}
\begin{align}
    \label{equ:V0}
    V(\Phi, X) &= \frac{\mu_2^2}{2}{\rm Tr}(\Phi^\dagger\Phi)+\frac{\mu_3^2}{2}{\rm Tr}(X^\dagger X) + \lambda_1[{\rm Tr}(\Phi^\dagger\Phi)]^2 + \lambda_2{\rm Tr}(\Phi^\dagger\Phi){\rm Tr}(X^\dagger X) \nonumber \\
    & \qquad + \lambda_3{\rm Tr}(X^\dagger X X^\dagger X) + \lambda_4[{\rm Tr}(X^\dagger X)]^2 - \lambda_5{\rm Tr}(\Phi^\dagger \tau^a\Phi\tau^b){\rm Tr}(X^\dagger t^a X t^b),
\end{align}
where the possible cubic terms in the general GM model are eliminated under the $Z_2$ symmetry. The SU(2) generators for doublet representation are $\tau^a=\sigma^a/2$, where $\sigma^a$ are the Pauli matrices. For the triplet, the generators are given by
\begin{align}
    t^1 = \frac{1}{\sqrt{2}}\begin{pmatrix}
        0 & 1 & 0 \\
        1 & 0 & 1 \\
        0 & 1 & 0
    \end{pmatrix}, \quad t^2=\frac{1}{\sqrt{2}}\begin{pmatrix}
        0 & -i & 0 \\
        i & 0 & -i \\
        0 & i & 0
    \end{pmatrix}, \quad t^3=\begin{pmatrix}
        1 & 0 & 0 \\
        0 & 0 & 0 \\
        0 & 0 & -1
    \end{pmatrix}.
\end{align}
In terms of the scalar VEV, the scalar potential can be expressed as
\begin{align}
    V_0 = \frac{\mu_2^2}{2}v_\phi^2 + \lambda_1 v_\phi^4.
\end{align}
Then the minimization condition is given by
\begin{align}
    \frac{\partial V_0}{\partial v_\phi} = \mu_{2}^{2}v_{\phi} + 4\lambda_{1}v_{\phi}^{3} = 0.
\end{align}

After electroweak symmetry breaking, the $SU(2)_L$ doublets and triplets can be decomposed into mass eigenstates which contain two custodial singlets ($h,H$), two custodial triplets ($G,H_3$), and one custodial fiveplet ($H_5$). One of the singlets ($h$) will be identified as the 125 GeV Higgs boson. One of the triplets will be the goldstone corresponding to the broken symmetries. Note that the scalars ($h,G$) from the $SU(2)_L$ doublet will be $Z_2$-even, while the scalars from the triplets ($H,H_3,H_5$) will be $Z_2$-odd.
It should be noted that the lightest neutral component of the $Z_2$-odd scalars will be the DM candidate. The mass spectrum of the scalars are given by
\begin{subequations}
\label{equ:masses}
\begin{align}
    m_h^2 &= 8 \lambda _1 v_\phi^2, \\
    m_H^2 &= \mu _3^2+(2 \lambda _2 -\lambda _5) v_\phi^2=\mu_3^2+\frac{1}{2} \lambda_{Hh}v_\phi^2, \\
    m_{3}^2 &=\mu _3^2+(2 \lambda _2-\frac{1}{2}\lambda _5) v_\phi^2=\mu_3^2+\frac{1}{2} \lambda_{H_3h}v_\phi^2, \\
    m_{5}^2 &=  \mu _3^2+(2 \lambda _2+ \frac{1}{2}\lambda _5 ) v_\phi^2=\mu_3^2+ \frac{1}{2}\lambda_{H_5h}v_\phi^2,
\end{align}
\end{subequations}
where $\lambda_{Sh}$ with $S=H,H_3,H_5$ represents the $SSh$ cubic coupling. If $\lambda_5>0$, that is, $\lambda_{Hh}<\lambda_{H_3h}<\lambda_{H_5h}$ in the tree-level parameterization, then $m_H<m_{3}< m_{5}$; otherwise, $m_H>m_{3}> m_{5}$.

For convenience, we use zero-temperature masses as inputs instead of the $\lambda_i$'s. Upon examining $m_H^2$, $m_{3}^2$, and $m_{5}^2$, we observe that although these masses are determined by four parameters, that is, $\mu_3^2$, $\lambda_2$, $\lambda_5$, $v_\phi$, they depend on these parameters only through two combinations: $\mu_3^2 + 2\lambda_2 v_\phi^2$ and $\lambda_5 v_\phi^2$. Consequently, if we wish to use those masses as input parameters, we can only select two of them, with the third determined by these two. In principle, any two of these masses can be chosen. However, since $m_{3}$ always lies between $m_H$ and $m_{5}$, it is more convenient to use $m_{H}$ and $m_{5}$ as input parameters. Therefore, we can treat $\lambda_{1}, \lambda_{5}, \mu_2^2, \mu_3^2, m_3^2$ as functions of $ v_\phi^2, m_h^2, m_H^2, m_5^2, \lambda_{2}$ (the zero-temperature physical quantities) as following
\begin{subequations}
    \begin{align}
        \lambda_1 &= \frac{m_h^2}{8 v_\phi^2}, \\
        \lambda_5 &= \frac{2(m_{5}^2-m_H^2)}{3 v_\phi^2}, \\
        \mu_{2}^2 & = -4 \lambda _1 v_\phi^2, \\
        \mu_{3}^2 &= \frac{m_H^2+2m_{5}^2-6\lambda_2 v_\phi^2}{3}, \\
        m_{3}^2 &=  \frac{m_{5}^2+2m_H^2}{3}.
    \end{align}
    \label{equ:lambda}
\end{subequations}
In addition, $\lambda_3$ and $\lambda_4$ are two free parameters which determine the interactions among $Z_2$-odd scalars.

\begin{table}
\centering
\begin{tabular}{|c|c|}
\hline\hline
Fixed Parameters & Scanned Parameters \\
\hline
$v_\phi = 246\, \rm GeV$ & $m_H\in[10, 2000]\,\rm GeV$ \\
$m_h = 125\, \rm GeV$ & $m_{5}\in[10, 2000]\,\rm GeV$ \\
 & $\lambda_2\in[-5, 5]$ \\
 & $\lambda_3\in[-5, 5]$ \\
 & $\lambda_4\in[-5, 5]$ \\
\hline\hline
\end{tabular}
\caption{The GM model parameters used in this work.}
\label{tab:parameters}
\end{table}

Hence, instead of the parameters in the scalar potential $(\mu_2^2,\mu_3^2,\lambda_1,\lambda_2,\lambda_3,\lambda_4,\lambda_5)$, we will use as many physical parameters as possible as free parameters:
\begin{align}
    v_\phi=246\,{\rm GeV}, m_h=125\,{\rm GeV}, m_H, m_5, \lambda_2, \lambda_3,\lambda_4.
\end{align}
The parameter choices are given in~\autoref{tab:parameters}. The parameter scan is performed using the code {\tt GMCalc}~\cite{Hartling:2014xma} with the following theoretical constraints included.

\subsection{Theoretical constraints}
\label{sec2_2}

Three theoretical constraints should be applied before we consider the physical results: tree-level unitarity, the bounded-from-below (BFB) condition on the potential, and the absence of deeper minima other than $v_\phi = 246, v_\chi = 0$. These constraints are generally implemented in terms of the potential parameters $\lambda_i$ (scalar quartic couplings)~\cite{Hartling:2014xma,Hartling:2014zca}.

The unitarity conditions are given by~\cite{Aoki:2007ah, Hartling:2014zca}
\begin{align}
\sqrt{\left(6 \lambda_{1}-7 \lambda_{3}-11 \lambda_{4}\right)^{2}+36 \lambda_{2}^{2}}+\left|6 \lambda_{1}+7 \lambda_{3}+11 \lambda_{4}\right| & <4 \pi, \nonumber\\
\sqrt{\left(2 \lambda_{1}+\lambda_{3}-2 \lambda_{4}\right)^{2}+\lambda_{5}^{2}}+\left|2 \lambda_{1}-\lambda_{3}+2 \lambda_{4}\right| & <4 \pi, \nonumber\\
\left|2 \lambda_{3}+\lambda_{4}\right| & <\pi, \nonumber \\
\left|\lambda_{2}-\lambda_{5}\right| & <2 \pi.
\end{align}
The BFB condition further requires that
\begin{equation}
\begin{array}{l}
\lambda_{1}>0, \\
\lambda_{4}>\left\{\begin{array}{ll}
-\frac{1}{3} \lambda_{3} \text { for } \lambda_{3} \geq 0, \\
-\lambda_{3} \text { for } \lambda_{3}<0,
\end{array}\right. \\
\lambda_{2}>\left\{\begin{array}{ll}
\frac{1}{2} \lambda_{5}-2 \sqrt{\lambda_{1}\left(\frac{1}{3} \lambda_{3}+\lambda_{4}\right)} & \text { for } \lambda_{5} \geq 0 \text { and } \lambda_{3} \geq 0, \\
\omega_{+}(\zeta) \lambda_{5}-2 \sqrt{\lambda_{1}\left(\zeta \lambda_{3}+\lambda_{4}\right)} & \text { for } \lambda_{5} \geq 0 \text { and } \lambda_{3}<0, \\
\omega_{-}(\zeta) \lambda_{5}-2 \sqrt{\lambda_{1}\left(\zeta \lambda_{3}+\lambda_{4}\right)} & \text { for } \lambda_{5}<0,
\end{array}\right.
\end{array}
\end{equation}
with
\begin{equation}
\omega_{ \pm}(\zeta)=\frac{1}{6}(1-B) \pm \frac{\sqrt{2}}{3}\left[(1-B)\left(\frac{1}{2}+B\right)\right]^{1 / 2}
\end{equation}
and
\begin{equation}
B \equiv \sqrt{\frac{3}{2}\left(\zeta-\frac{1}{3}\right)} \in[0,1] ; \quad \zeta \in\left[\frac{1}{3}, 1\right].
\end{equation}
Finally, we need to ensure that the dark matter vacuum ($v_\phi = 246, v_\chi = 0$) is the global minimum of the scalar potential. This condition is checked numerically in the code {\tt GMCalc}~\cite{Hartling:2014xma}.

\subsection{Electroweak Phase Transition in $Z_2$ symmetric GM model}
\label{sec:EWPT}
The model may occupy different vacuum states during the cooling of the universe. Transiting from one state to another is proceeded through phase transition. For simplicity, in this work, we only consider the vacuum states that preserve the custodial symmetry
\begin{align}
    \langle\Phi\rangle = \frac{\omega_\phi}{\sqrt{2}}\mathbb{I}_{2\times 2},\quad \langle X\rangle = \omega_\chi\mathbb{I}_{3\times 3},
\end{align}
where $\omega_\phi|_{T=0}=v_\phi=246\,\rm GeV$, $\omega_\chi|_{T=0}=v_\chi=0$. Then the tree-level zero temperature effective potential reads
\begin{align}
V_0=V(\omega_\phi,\omega_\chi)&=\frac{\mu_2^2}{2}\omega_\phi^2 + \frac{3\mu_3^2}{2}\omega_\chi^2 + \lambda_1 \omega_\phi^4 + 3(\lambda_3+3\lambda_4)\omega_\chi^4 + \frac{3}{2}\left(2\lambda_2 - \lambda_5\right)\omega_\phi^2\omega_\chi^2 \nonumber \\
&=\frac{\mu_2^2}{2}\omega_\phi^2 + \frac{3\mu_3^2}{2}\omega_\chi^2 + \lambda_1 \omega_\phi^4 + 3\lambda_{34}\omega_\chi^4 + \frac{3}{2}\lambda_{25}\omega_\phi^2\omega_\chi^2,
\end{align}
where $\lambda_{34}\equiv \lambda_3 + 3\lambda_4$ and $\lambda_{25}\equiv2\lambda_2 - \lambda_5$.
The effective potential receives higher-order corrections. In this work, we consider high temperature expansion of the effective potential, which have the advantage of being gauge invariant~\cite{Arnold:1992rz,Bian:2019kmg} and can be expressed as
\begin{align}
\label{equ:Veff}
V_{\rm eff} &= V_0 + V_{1,T} = V_0 + \frac{1}{2}c_2T^2\omega_\phi^2 + \frac{1}{2}c_3T^2\omega_\chi^2,
\end{align}
with
\begin{align}
    c_2 = \frac{3g^2+g'^2}{16} + \frac{y_t^2}{4} + 2\lambda_1 + \frac{3}{2}\lambda_2,\quad c_3 = \frac{3g^2+g'^2}{2} + 2\lambda_2 + 7\lambda_3 + 11\lambda_4.
\end{align}

In order to determine the vacuum structure, we need to find the field values that extremize the $V_{\rm eff}$, and the tadpole equations are
\begin{align}
    \frac{\partial V_{\rm eff}}{\omega_\phi}&=\omega_\phi(\tilde{\mu}_2^2+4\lambda_1\omega_\phi^2+3\lambda_{25}\omega_\chi^2)=0, \\
    \frac{\partial V_{\rm eff}}{\omega_\chi}&=3\omega_\chi(\tilde{\mu}_3^2+4\lambda_{34}\omega_\chi^2+\lambda_{25}\omega_\phi^2)=0,
\end{align}
where $\tilde{\mu}_2^2\equiv \mu_2^2 + c_2T^2$ and $\tilde{\mu}_3^2\equiv \mu_3^2 + \frac{1}{3}c_3T^2$.
Then the model has only 4 possible vacua solutions
\begin{subequations}
\label{equ:all_vacua}
\begin{align}
    1.&\quad \omega_\phi=0, \quad &\omega_\chi&=0;\\
    2.&\quad \omega_\phi=0, \quad &\omega_\chi^2&=-\frac{\tilde{\mu}_3^2}{4\lambda_{34}};
    \label{equ:two}\\
    3.&\quad \omega_\phi^2=-\frac{\tilde{\mu}_2^2}{4\lambda_{1}}, \quad &\omega_\chi&=0;
    \label{equ:three}\\
    4.&\quad \omega_\phi^2=\frac{3 \lambda _{25} \tilde{\mu}_3^2-4 \lambda _{34} \tilde{\mu}_2^2}{16 \lambda _1 \lambda _{34}-3 \lambda _{25}^2}, \quad &\omega_\chi^2&=\frac{\lambda _{25} \tilde{\mu} _2^2-4 \lambda _1 \tilde{\mu} _3^2}{16 \lambda _1 \lambda _{34}-3 \lambda _{25}^2}.
    \label{equ:four}
\end{align}
\end{subequations}
In order to ascertain whether the aforementioned solutions can represent local minima of the potential, it is necessary to compute the local minimum conditions corresponding to each solution. At the extremum points, the second order derivatives of the effective potential can be assembled into a Hessian matrix. The field and temperature-dependent Hessian matrix~\cite{Ghorbani:2020xqv} is given by
\begin{equation}
\mathcal{H}=\begin{pmatrix}
\tilde{\mu}_2^2+12 \lambda _1 \omega_{\phi}^2+3 \lambda _{25} \omega_\chi^2 & 6 \lambda _{25} \omega_{\phi} \omega_\chi\\
6 \lambda _{25} \omega_{\phi} \omega_\chi & 3 \tilde{\mu} _3^2+3 \lambda _{25} \omega_{\phi}^2+36 \lambda _{34} \omega_\chi^2
\end{pmatrix}.
\label{equ:ori-Hessian}
\end{equation}
The Hessian matrix has to be positive-defined for local minima. Hence it is straightforward that only vacuum-2 and vacuum-3 in~\autoref{equ:all_vacua} can be local minima simultaneously.

At zero temperature, we assume that the system is in the $Z_2$ phase $(\omega_\phi = v_\phi = 246\,{\rm GeV},\, \omega_\chi =v_\chi= 0)|_{T=0}$ in order to have stable DM candidate. For this case, the Hessian matrix is given as
\begin{equation}
\mathcal{H}=\begin{pmatrix}
 -2\mu _2^2 & 0\\
0 &\frac{ 12\lambda_{1} \mu _3^2-3 \lambda _{25}\mu_2^2}{4\lambda_{1}}
\end{pmatrix},
\end{equation}
which requires
\begin{equation}
\mu_2^2=-4\lambda_1v_\phi^2<0,\quad \lambda_1>0,\quad 4\lambda_1\mu_3^2>\lambda_{25}\mu_2^2.
\end{equation}
In addition, in order to ensure $(v_\phi,0)$ as a global minimum at $T=0$, we need to have (even if $(0,v_\chi)$ is a saddle point)
\begin{equation}
\label{equ:T0_condition}
-\frac{\mu_2^4}{16\lambda_1}<-\frac{3\mu_3^4}{16\lambda_{34}}\quad \Rightarrow\quad \mu_3^2\sqrt{3\lambda_1}>\mu_2^2\sqrt{\lambda_{34}}.
\end{equation}

Furthermore, in our case for the effective potential given in~\autoref{equ:Veff} with $Z_2$ phase at $T=0$, it can be proved that the only way to achieve first-order phase transition is through the second step in a two step transition of $(0,0)\to (0,\omega_\chi)\to (\omega_\phi,0)$. The necessary conditions for SFOEWPT are thus the existence of the two phases $(0,\omega_\chi)$, $(\omega_\phi,0)$, and these two phases are degenerate at critical temperature $T_c$, which requires
\begin{align}
\label{equ:h_Tc}
&\mu_2^2+c_2T_c^2<0,\ 4\lambda_1(3\mu_3^2+c_3T_c^2)>3\lambda_{25}(\mu_2^2+c_2T_c^2),\ \omega_\phi^2(T_c)=-\frac{\mu_2^2+c_2T_c^2}{4\lambda_1};\\
\label{equ:s_Tc}
&3\mu_3^2+c_3T_c^2<0,\ 4\lambda_{34}(\mu_2^2+c_2T_c^2)>\lambda_{25}(3\mu_3^2+c_3T_c^2),\ \omega_\chi^2(T_c)=-\frac{3\mu_3^2+c_3T_c^2}{12\lambda_{34}},
\end{align}
and
\begin{equation}
V(\omega_\phi(T_c),0)=-\frac{(\mu_2^2+c_2T_c^2)^2}{16\lambda_1}=V(0,\omega_\chi(T_c))=-\frac{(3\mu_3^2+c_3T_c^2)^2}{48\lambda_{34}}.
\end{equation}
Solving the above equation yields
\begin{equation}
T_c^2=\frac{\sqrt{3\lambda_{34}}\mu_2^2 - 3\sqrt{\lambda_1}\mu_3^2}{c_3\sqrt{\lambda_1}-c_2\sqrt{3\lambda_{34}}}.
\end{equation}
Combining with the condition given in~\autoref{equ:T0_condition}, we have
\begin{align}
\label{equ:Tc_condition}
    c_3\sqrt{\lambda_1} < c_2\sqrt{3\lambda_{34}}.
\end{align}
Substituting the expression of $T_c$ into~\autoref{equ:h_Tc} and~\autoref{equ:s_Tc} and combining with~\autoref{equ:Tc_condition}, one obtains
\begin{align}
 &c_3\mu_2^2>3c_2\mu_3^2, \quad 4\lambda_1\sqrt{3\lambda_{34}} < 3\lambda_{25}\sqrt{\lambda_1}.
\end{align}
Finally, by combining all the conditions listed above, we reach the necessary condition for first-order EWPT in $Z_2$ symmetric GM model, which is
\begin{equation}
\label{equ:SFOEWPT_Conditions}
\frac{c_3}{c_2}=\frac{\frac{3g^2+g'^2}{2}+2\lambda_2+7\lambda _3+11\lambda _4}{\frac{3g^2+g'^2}{16}+\frac{y_t^2}{4}+2\lambda_1+\frac{3}{2}\lambda _2}<\frac{3\mu_3^2}{\mu_2^2}<\frac{\sqrt{3\lambda_{34}}}{\sqrt{\lambda_1}}<\frac{3\lambda_{25}}{4\lambda_1}.
\end{equation}

Strong first-order EWPT (SFOEWPT) can induce detectable gravitational wave signal. The transition strength parameter $\alpha$ and the ratio of the inverse time duration of transition to the Hubble constant $\frac{\beta}{H_ {*}}$ are two key parameters in the phase transition determining the frequency and strength of the GW signal. The $\alpha$ and $\frac{\beta}{H_ {*}}$ are defined as
\begin{equation}
\alpha=\frac{\Delta \rho}{\rho_{R}},\quad \frac{\beta}{H_{*}}=\left.T\frac{ dS}{dT}\right|_{T=T_{*}},
\end{equation}
where $\rho_{R}=\frac{\pi^2}{30}g_{*}T_{*}^4$ is the radiation energy density ($g_{*}$ is the effective number of relativistic degrees of freedom at the temperature $T_{*}$), $S=\frac{S_{3}}{T}$ ($S_3$ is the Euclidean action of the $O(3)$-symmetric bubble~\cite{Linde:1981zj}) is the bounce action, $\Delta \rho$
is the difference of energy density between the false and true vacua, which can be obtained by~\cite{Enqvist:1991xw}
\begin{align}
    \Delta \rho&=\rho(T_*)|_{(0,\omega_\chi)}-\rho(T_*)|_{(\omega_\phi,0)}, \\
    \rho(T)&=V_{\rm eff}\left(T\right)-T \frac{\partial V_{\rm eff}\left(T\right)}{\partial T}.
\end{align}
We implemented the $Z_2$ symmetric GM model in {\tt CosmoTransitions}~\cite{Wainwright:2011kj} for the calculation of the relevant parameter of EWPT.

The GW from EWPT mainly originated from three sources: bubble collisions $\Omega_{\phi}$, sound waves $\Omega_{sw}$, and Magnetohydrodynamic (MHD) turbulence $\Omega_{turb}$.
Based on the assumption of non-runaway bubbles~\cite{Caprini:2015zlo}, that is, the velocity of the bubble wall is affected by plasma friction and cannot reach ultra-relativistic speeds, the energy contribution of bubble wall collisions to gravitational waves is negligible. Therefore, the gravitational wave spectrum is mainly determined by the contributions of sound waves and turbulence, which can be expressed as
\begin{align}
    h^2\Omega_{GW}=h^2\Omega_{sw}+h^2\Omega_{turb},
\end{align}
with
\begin{align}
        h^{2} \Omega_{\mathrm{sw}}(f)&=2.65 \times 10^{-6}\left(\frac{H_{*}}{\beta}\right)\left(\frac{\kappa_{v} \alpha}{1+\alpha}\right)^{2}\left(\frac{100}{g_{*}}\right)^{\frac{1}{3}} v_{w} S_{\mathrm{sw}}(f), \\
        h^{2} \Omega_{\mathrm{turb}}(f)&=3.35 \times 10^{-4}\left(\frac{H_{*}}{\beta}\right)\left(\frac{\kappa_{\mathrm{turb}} \alpha}{1+\alpha}\right)^{\frac{3}{2}}\left(\frac{100}{g_{*}}\right)^{1 / 3} v_{w} S_{\mathrm{turb}}(f),
\end{align}
where $\kappa_{v}$ is the fraction of latent heat transformed into the plasma, and $\kappa_{\mathrm{turb}}$ is the fraction of latent heat transferred to MHD~\cite{Espinosa:2010hh}, which can be written in the form of
\begin{equation}
   \kappa_{v}=\frac{\sqrt{\alpha}}{0.135+\sqrt{0.98+\alpha}} ,\quad \kappa_{\mathrm{turb}}=0.05\kappa_{v}\sim 0.1\kappa_{v}.
\end{equation}
Moreover, $S_{\mathrm{sw}}(f)$ and $S_{\mathrm{turb}}(f)$ parameterize the spectral shape of gravitational wave radiation and are given by~\cite{Caprini:2015zlo}
\begin{align}
    S_{\mathrm{sw}}(f)&=\left(f / f_{\mathrm{sw}}\right)^{3}\left(\frac{7}{4+3\left(f / f_{\mathrm{sw}}\right)^{2}}\right)^{7 / 2}, \, \\
    S_{\mathrm{turb}}\left(f\right)&=\frac{\left(f/f_{\mathrm{turb}}\right)^{3}}{\left[1+\left(f/f_{\mathrm{turb}}\right)\right]^{\frac{11}{3}}\left(1+8\pi f/h_{*}\right)}.
\end{align}
The peak frequencies after redshift are
\begin{align}
f_\mathrm{sw}&=1.9\times10^{-2}\mathrm{mHz}\frac{1}{v_w}\left(\frac{\beta}{H_*}\right)\left(\frac{T_*}{100\mathrm{GeV}}\right)\left(\frac{g_*}{100}\right)^{\frac{1}{6}},\\
f_{\mathrm{turb}} &=2.7\times10^{-2}\mathrm{mHz}\frac{1}{v_{w}}\left(\frac{\beta}{H_{*}}\right)\left(\frac{T_{*}}{100\mathrm{GeV}}\right)\left(\frac{g_{*}}{100}\right)^{\frac{1}{6}},
\end{align}
where $v_{w}$ represents the velocity of the bubble wall. In this work, we adopt
$v_{\omega}=\frac{\sqrt{1 / 3}+\sqrt{\alpha^{2}+2 \alpha / 3}}{1+\alpha}$~\cite{Espinosa:2010hh}.

\section{DM relic abundance in $Z_2$ symmetric GM model}
\label{sec:relic_density}
In $Z_2$ symmetric GM model,
the scalars coming from the $SU(2)_L$ triplet, including the custodial triplet, fiveplet and the extra singlet, will be odd under the $Z_2$ symmetry.
The corresponding neutral component will be natural candidate of the DM\footnote{The corresponding charged components will have a slightly higher mass due to radiation corrections~\cite{Cirelli:2005uq}.}. However, in the $Z_2$ symmetric case, only $H$ or $H_5^0$ can be the lightest one according to~\autoref{equ:masses}. Hence, we will discuss the situations where $H$ (case-$H$) or $H_5^0$ (case-$H_5$) is the DM candidate separately.

\begin{figure}[!tb]
    \centering
    \includegraphics[width=0.32\textwidth]{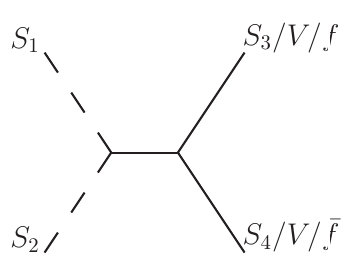}
    \includegraphics[width=0.32\textwidth]{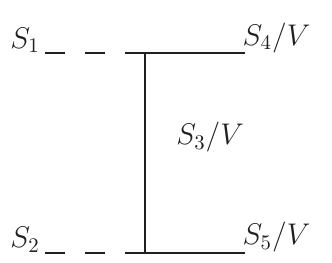}
    \includegraphics[width=0.32\textwidth]{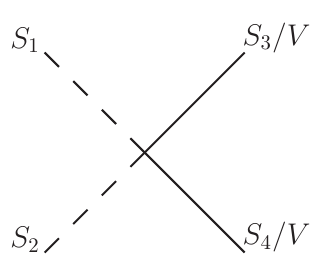}
    \caption{Tree-level Feynman diagrams for the self-annihilation and co-annihilation of DM into BSM scalars ($S_i\in\{H, H_3^0, H_3^\pm, H_5^0, H_5^\pm, H_5^{\pm\pm}\}$), gauge bosons ($W^\pm$), and fermions ($b$, $c$, $\tau^-$). Diagrams corresponding to u-channel processes are not shown separately.}
    \label{fig:feynman_diagrams}
\end{figure}

\begin{table}[!tbp]
\centering
\begin{tabular}{l|l|l}
\hline\hline
Processes & Proc-L & Proc-H \\
\hline
case-$H$ & $HH\to WW$ & $HH\to H_3^+H_3^-$ \\
& $HH\to ZZ$ & $HH\to H_5^+H_5^-,\, H_5^{++}H_5^{--}$ \\
& $HH\to hh$ & $HH_3^0\to f\bar{f},\, Zh,\, S_iS_j$\\
& $HH\to t\bar{t},\, b\bar{b},\, c\bar{c},\,\tau^+\tau^-$ & $HH_5^0\to ZZ,\, S_iS_j$ \\
&  & $H_3^0H_3^{0}\to f\bar{f},\, hh,\, ZZ,\, W W,\, S_iS_j$ \\
&  & $H_3^0H_5^{0}\to f\bar{f},\,  S_iS_j$ \\
&  & $H_5^0H_5^0\to hh,\, ZZ,\, W W,\, S_iS_j$ \\
\hline
case-$H_5$ & $H_5^0H_5^0\to WW$ & $H_5^0H_5^0\to H_3^\pm H_5^\mp, H_3^+H_3^-$ \\
 & $H_5^0H_5^0\to ZZ$ & $H_5^0H_3^0\to f\bar{f},\, S_iS_j$ \\
 & $H_5^0H_5^0\to hh$ & $H_5^0H\to  ZZ,\, S_iS_j$ \\
 & $H_5^0H_5^0\to t\bar{t},\, b\bar{b},\, c\bar{c},\,\tau^+\tau^-$ & $H_3^0H_3^0\to  hh,\, ZZ,\, W W,\, S_iS_j$ \\
 & $H_5^0H_5^0\to H_5^{++}H_5^{--},\,H_5^+H_5^-$ & $H_3^0H\to f\bar{f},\, S_iS_j$ \\
 &  & $HH\to hh,\, ZZ,\, W W,\, S_iS_j$ \\
\hline\hline
\end{tabular}
\caption{Major processes contributing to the DM relic density for both case-$H$ and case-$H_5$ grouped according to the particles involved in the process other than the DM particle , where $S_i$ and $S_j$ belong to the set of BSM scalars $\left\{H, H_3^0, H_3^\pm, H_5^0, H_5^\pm, H_5^{\pm\pm}\right\}$.}
\label{tab:processes}
\end{table}

\begin{figure}[!btp]
\centering
\includegraphics[width=0.48\textwidth]{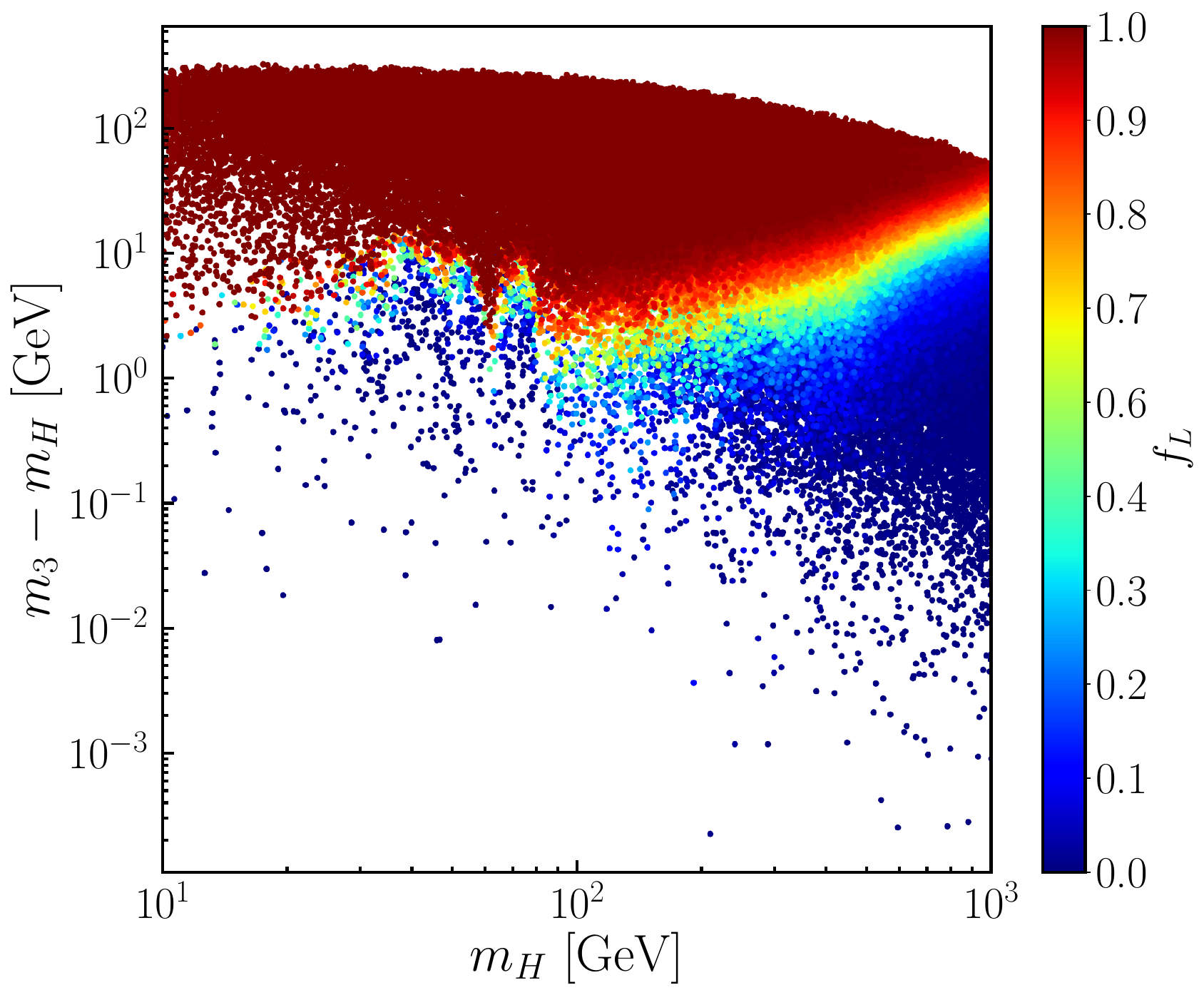}
\includegraphics[width=0.48\textwidth]{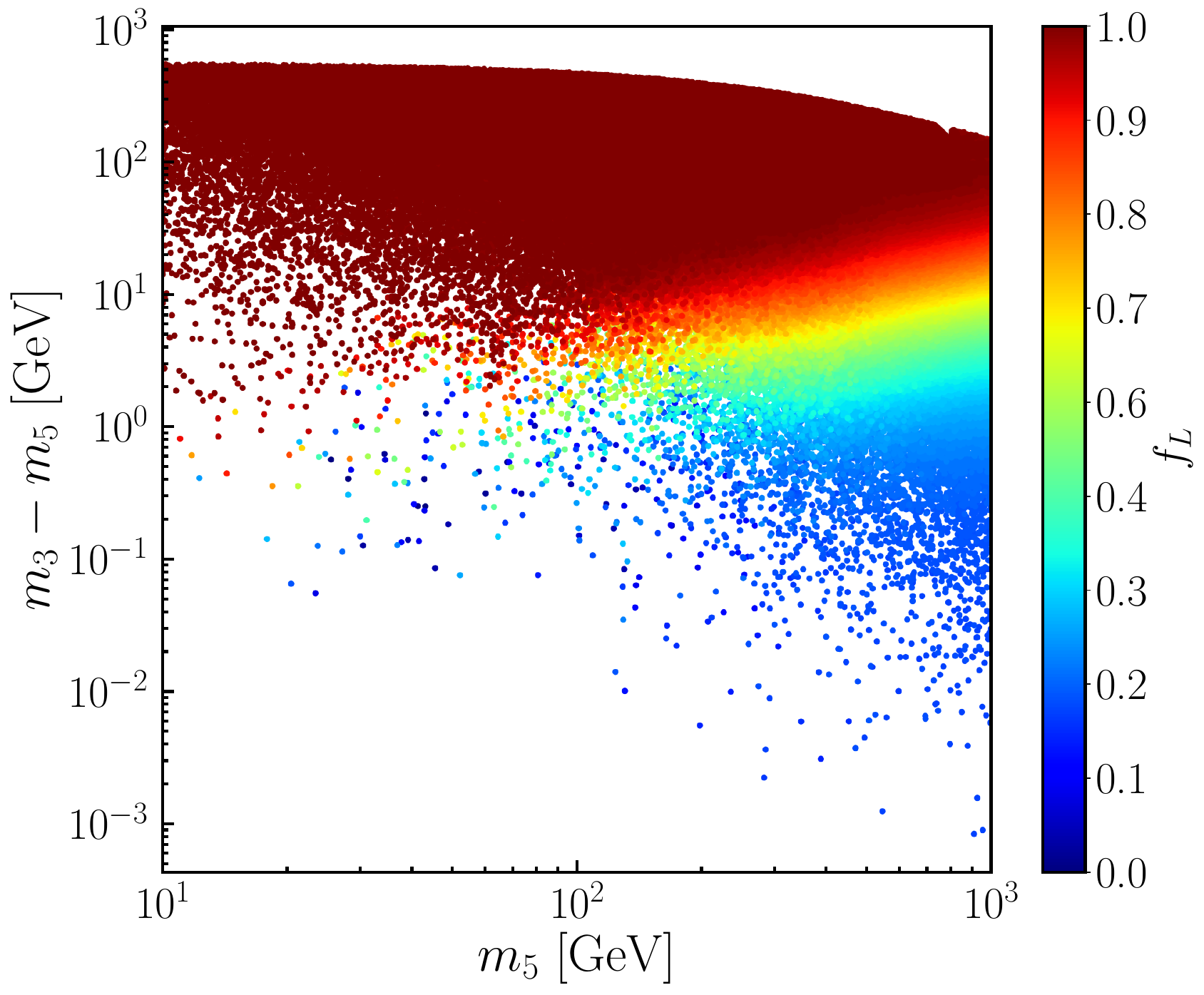}
\caption{\label{fig:delta-M-fl}
The relative contribution to the final DM relic density from the processes in {\it Proc-L} $f_L$ for case-$H$ (left panel) and case-$H_5$ (right panel). The color of the points indicates the relative contribution to the final DM relic density from the processes in {\it Proc-L}.
}
\end{figure}

Relevant processes for the DM annihilation are shown in~\autoref{fig:feynman_diagrams} including $s$-channel (left panel) where the mediator can be SM Higgs boson or massive gauge bosons, $t/u$-channel (middle panel) where the $t/u$-channel mediator can be either scalars or massive gauge bosons, and contact interactions (right panel). For later convenience, the processes are grouped according to the particles involved in the process other than the DM particles.
\begin{description}
    \item[Proc-L]: Processes involving SM particles or other scalars in the same multiplet;
    \item[Proc-H]: Processes involving other heavier scalars.
\end{description}
 Major processes in above categories contributing to the DM relic density for both case-$H$ and case-$H_5$ are listed in~\autoref{tab:processes}. For case-$H$, the processes in {\it Proc-L} involve the trilinear coupling $g_{HHh}$ between the DM candidate $H$ and the SM Higgs through the Higgs mediated $s$-channel processes and the gauge couplings through the scalars mediated $t/u$-channel and contact interaction processes. For processes involving SM fermions in the final states, the Yukawa couplings are also involved. The cross-section of processes in {\it Proc-L} are hence mainly sensitive to the DM mass and the trilinear couplings $g_{HHh}$. On the other hand, the processes in {\it Proc-H} involve heavier scalars, contain mainly the co-annihilation processes. Although, these processes also depend on the gauge coupling and the trilinear couplings among scalars, the mass splittings among scalars are the dominant factors when considering the relic density of the DM which is mainly determined at the temperature well below the mass of the DM. The effects of mass splittings are shown in~\autoref{fig:delta-M-fl} where we define $f_L$ as the contribution to the final relic density from processes in {\it Proc-L}. It is clear that the processes in {\it Proc-H} can be dominant only when the mass splittings are relatively small.

For case-$H_5$, the situation is similar. However, the processes in {\it Proc-L} also involve those with other fiveplet scalars. Hence, the couplings among fiveplets which depend on $\lambda_{3,4}$ are also involved. From above naive prospects, the most important parameters are the DM mass, the mass splittings among scalars, the trilinear couplings $g_{H_5H_5h}$ between the DM candidates and the SM Higgs and quartic couplings among scalars. The Feynman rules for relevant interactions are listed in~\autoref{app:fr}.

\begin{figure}[!tbp]
\centering
\includegraphics[width=0.48\textwidth]{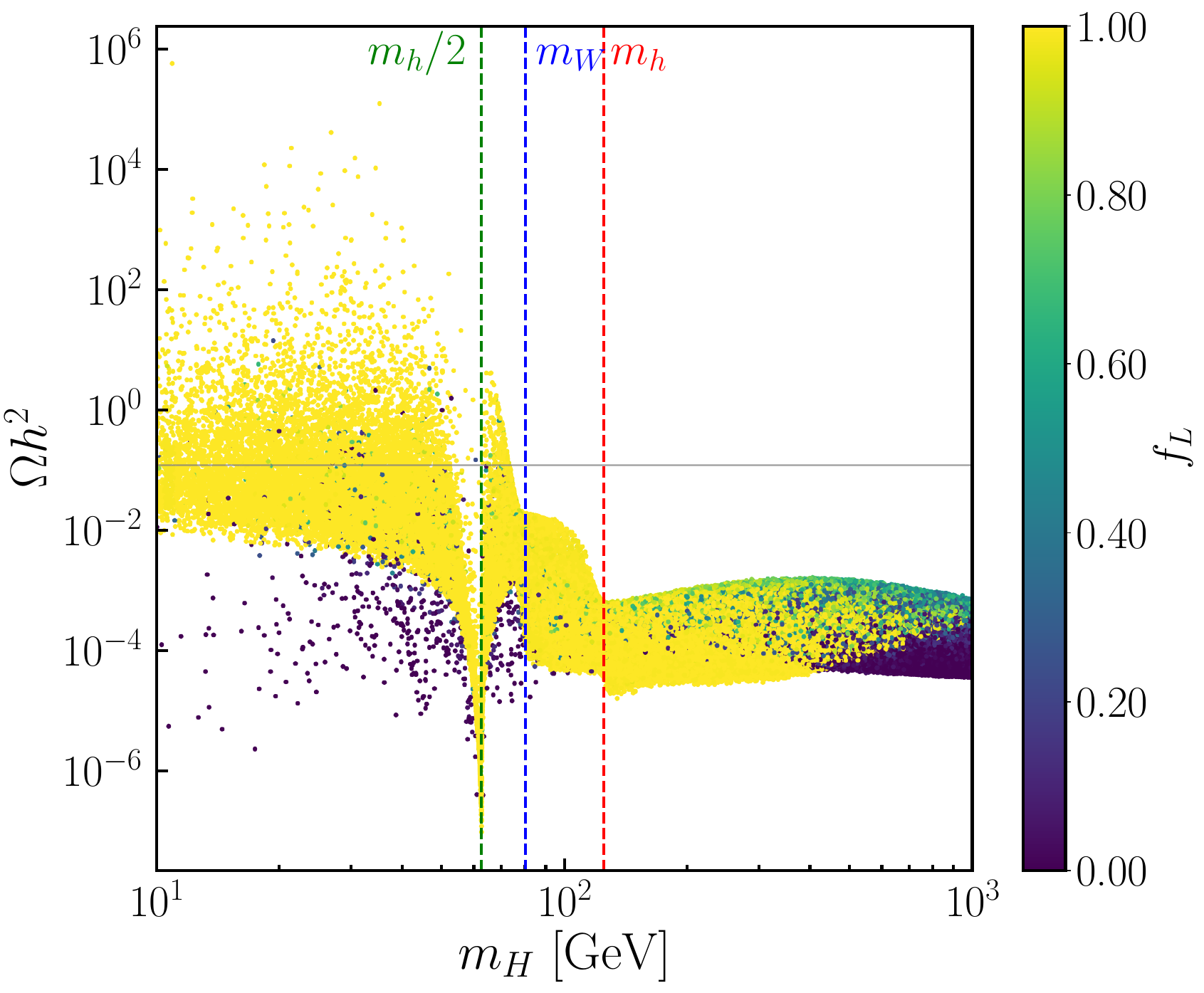}
\includegraphics[width=0.48\textwidth]{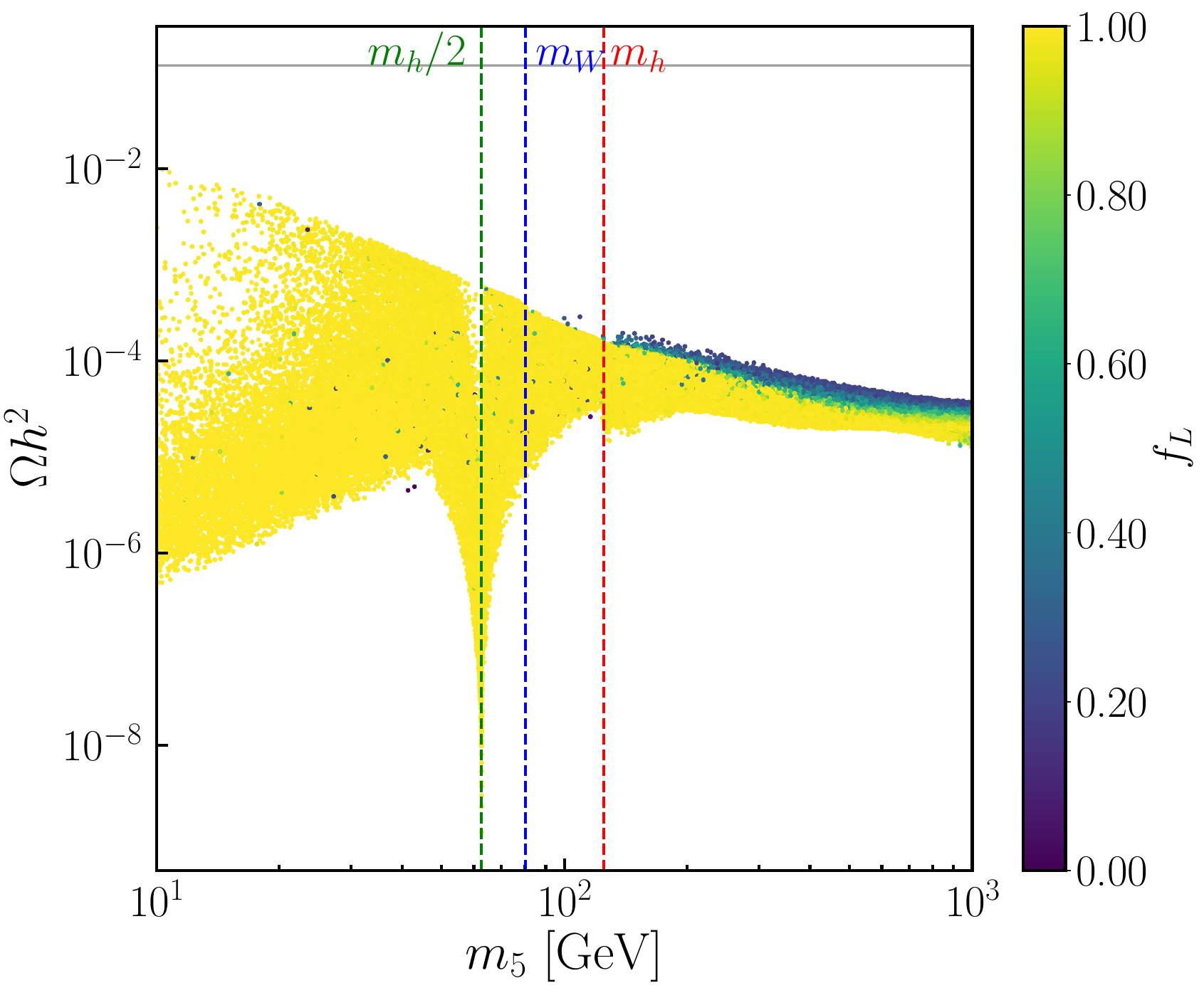}
\caption{The relic density of DM candidates as a function of the DM mass for case-$H$ (left) and case-$H_5$ (right). The gray horizontal band represents 2-$\sigma$ region around the current observed DM relic density. The blue, green and red vertical dashed lines represent $m_h/2$, $m_W$ and $m_h$, respectively. The color of the points indicates the relative contribution to the final DM relic density from the processes in {\it Proc-L}.
}
\label{fig:Omega-M}
\end{figure}

The predicted DM relic density, obtained from {\tt micrOMEGAs}~\cite{Alguero:2023zol}, as a function of the DM mass for both case-$H$ (left panels) and case-$H_5$ (right panels), is shown in~\autoref{fig:Omega-M}, where the horizontal gray line indicates the recent PLANCK measurement~\cite{Planck:2018vyg}, which is
\begin{equation}
\Omega_{\rm CDM}h^2=0.1200\pm 0.0012.
\label{equ:CDM}
\end{equation}
From the left panel of~\autoref{fig:Omega-M}, we can see that the relic density of the DM candidate in case-$H$ can be above, around (within 2-$\sigma$) or below the PLANCK measurement when the DM mass is less than the $W$ boson mass ($m_W$) which is indicated by the vertical blue dashed line. When the DM mass is large, the relic density of $H$ is always below the observation. By contrast, the relic density of the DM candidate in case-$H_5$ is always below the PLANCK measurement throughout the whole mass range as shown in the right panel of~\autoref{fig:Omega-M}. In the plots, we also indicate the half of the Higgs mass ($m_h/2$) by the vertical green dashed line, where the Higgs mediated $s$-channel processes dominate, and the relic density of the DM candidate is heavily suppressed.
Additionally, we mark the Higgs mass (\(m_h\)) with a vertical red dashed line. At this mass, DM particle annihilation into a Higgs boson pair becomes kinematically allowed, thereby suppressing the relic density of the DM candidate, as shown in the left panel of \autoref{fig:Omega-M}.

In~\autoref{fig:Omega-M}, we also show the relative contribution to the final relic density from the processes in {\it Proc-L} $f_L$ by the color of the points.
The processes that dominate the relic density are different for case-$H$ and case-$H_5$. For case-$H$, the final states in the {\it Proc-L} are only SM particles, the upper left panel of~\autoref{fig:Omega-M} clearly shows the threshold of $HH\to WW/ZZ$ and $HH\to hh$. While, on the other hand, there is no such pattern in the right panel of~\autoref{fig:Omega-M} as there are also processes involving other scalar in the fiveplet in {\it Proc-L} for case-$H_5$ as listed in~\autoref{tab:processes} that can dominate the relic density of DM.
Furthermore, for the same reason, the impact of the processes in {\it Proc-H} on the relic density is generally smaller in case-\(H_5\) than in case-\(H\), because the minimum value of \(f_L\) is larger in case-\(H_5\).

Since the parameters that determine the relic density of the DM candidate differ for processes in {\it Proc-L} and {\it Proc-H}, we will discuss separately the effects of the relevant parameters on the DM relic density, based on the relative contributions from the processes in {\it Proc-L} and {\it Proc-H}, in the following two scenarios:
\begin{description}
    \item[L-Dominant]: the relative contribution from the processes in {\it Proc-L} is larger than 60\%, $f_L > 0.6$.
    \item[Mixed]\footnote{Note that actually this also contains the situation where processes in {\it Proc-H} dominate.}: the relative contribution from the processes in {\it Proc-L} is less than 60\%, $f_L < 0.6$.
\end{description}

\subsection{case-$H$}

\begin{figure}[!tbp]
    \centering
    \includegraphics[width=0.48\linewidth]{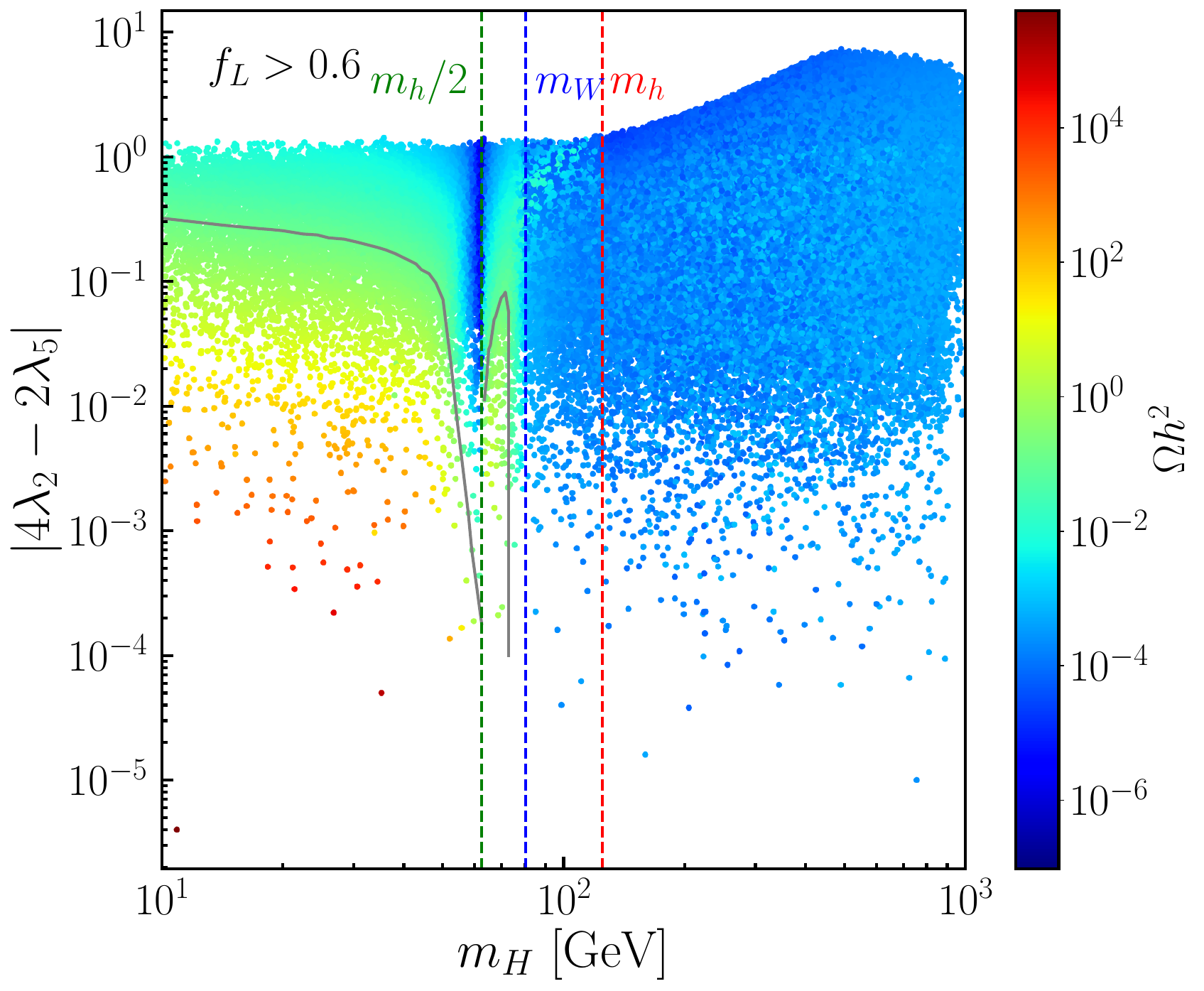}
    \caption{The relic density of the DM candidate for case-$H$ in the $|g_{HHh}/v_\phi|=|4\lambda_2-2\lambda_5|$ vs. $m_H$ plane for the {\it L-Dominant} scenario. The color of each point indicates the relic density and gray lines indicate the parameter region that matches the observed value for DM relic density. }
    \label{fig:case-H-relic-lam}
\end{figure}

For case-$H$, the processes in {\it Proc-L} are simple involving only self-annihilation of $H$ into SM particles. For fixed DM mass $m_H$, the cross section of these processes depends mainly on $g_{HHh}/v_\phi = 4\lambda_2 - 2\lambda_5$ through the Higgs mediated $s$-channel process. The relic density of the DM for {\it L-Dominant} scenario hence has straightforward dependence on these two parameters as shown in ~\autoref{fig:case-H-relic-lam}.
When the $m_H\lesssim m_W$, the processes with $W$ in the final states, which solely depend on the gauge coupling, are suppressed. In this low mass region, the Higgs mediated SM fermion pair production dominates, the region involving Higgs resonant effect can be clearly identified. In this case, a proper choice of the couplings $g_{HHh}/v_\phi = 4\lambda_2 - 2\lambda_5$ can achieve the observed relic density as shown by the thin gray lines in~\autoref{fig:case-H-relic-lam}. On the other hand, when $m_H\gtrsim m_W$, the processes with $W$ in the final states become important. The relic density of the DM candidate is hence heavily suppressed. The dependence on $g_{HHh}$ reduces in this region.

\begin{figure}[!tbp]
\centering
\includegraphics[width=0.48\textwidth]{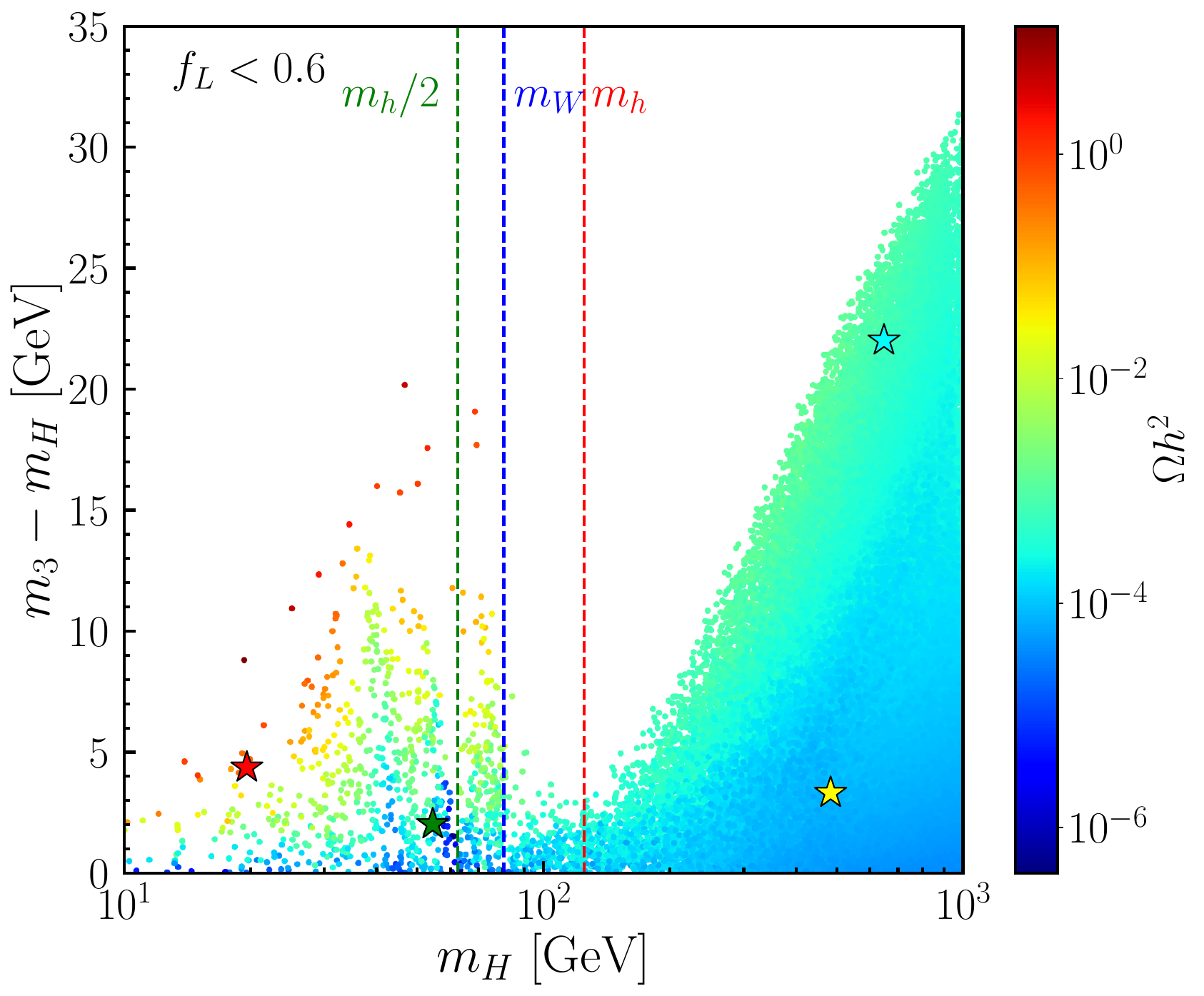}
\includegraphics[width=0.48\textwidth]{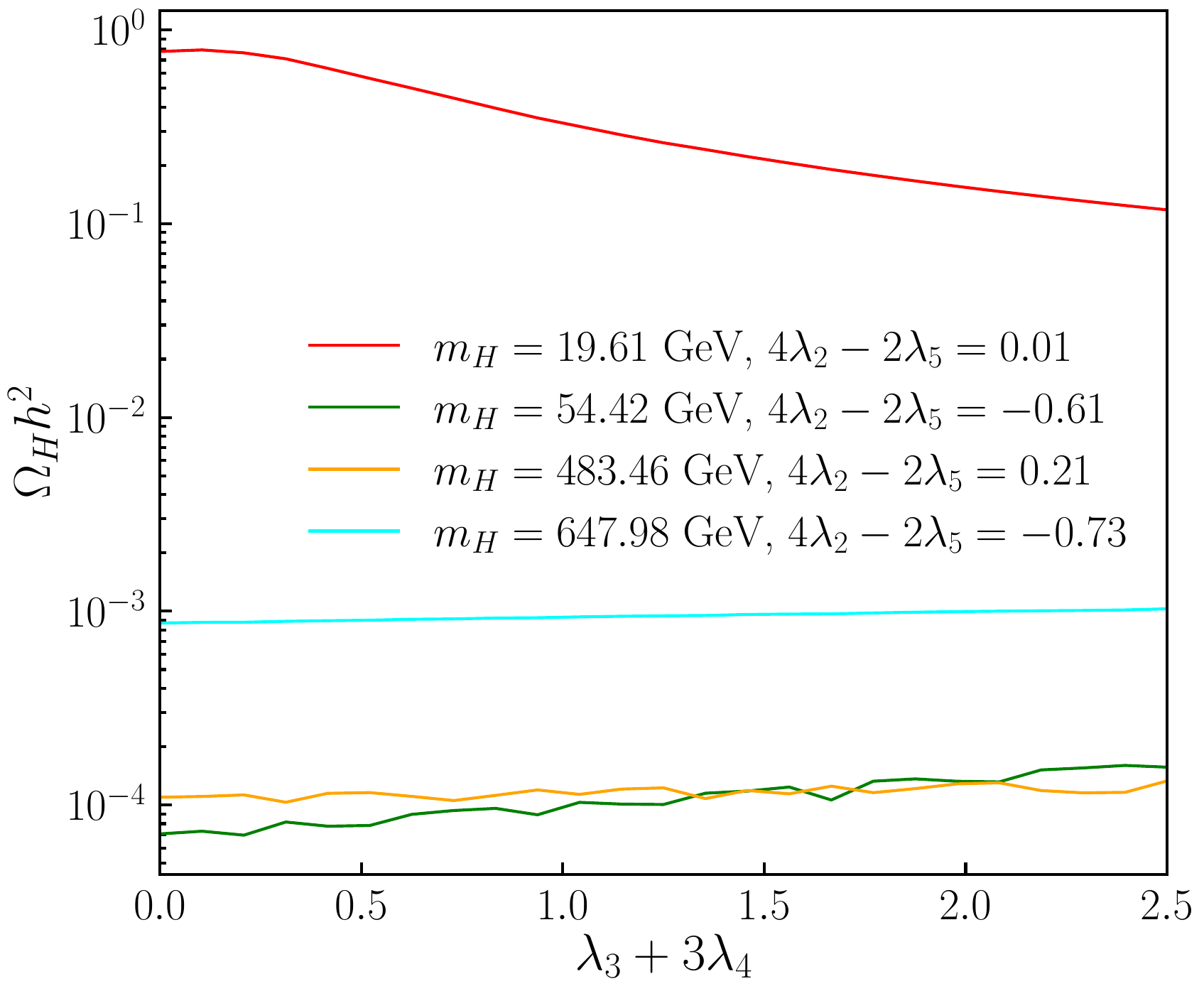}
\caption{For case-$H$, the DM relic density as a function of $\lambda_3+3\lambda_4$ in {\it Mixed} scenario.
Different colors represent different values of $g_{HHh}\sim4\lambda_2-2\lambda_5$.
}
\label{fig:case-H-relic-dm-lam3_4}
\end{figure}

On the other hand, for the {\it Mixed} scenario, the contribution from processes in {\it Proc-H} become important. This brings the dependence on the mass splitting and other quartic couplings among scalars.
As we have discussed above, these processes in {\it Mixed} scenario are important only when the mass splitting is relatively small. In the right panel of~\autoref{fig:case-H-relic-dm-lam3_4}, we show relic density in the $\Delta=m_3-m_H$ vs. $m_H$ plane.

When processes in {\it Proc-H} become important, the relic density will also depend on the quartic couplings among $Z_2$-odd scalars which depend on $\lambda_3$, $\lambda_4$. In particular, since the next-to-lightest scalars are always the triplet, the most important processes in {\it Proc-H} are then those containing only $H$ and $H_3$ which involve the quartic coupling $g_{HHH_3H_3}\sim\lambda_3+3\lambda_4$. For illustration of the dependence of the relic density on this quartic coupling, we pick several parameter points in the {\it Mixed} scenario with different values for $g_{HHh}\sim 4\lambda_2-2\lambda_5$ and $m_H$, and then show the DM relic density as function of $\lambda_3+3\lambda_4$ by scanning $\lambda_3$ and $\lambda_4$ randomly for these points in the right panel of~\autoref{fig:case-H-relic-dm-lam3_4}.
As clearly shown in~\autoref{fig:case-H-relic-dm-lam3_4}, with other parameters fixed, the DM relic density has a much simpler dependence on $\lambda_3+3\lambda_4$.
However, the dependence becomes more pronounced when $g_{HHh}\sim 4\lambda_2 - 2\lambda_5$ and $m_H$ are both small. In this scenario, the annihilation of the $W$ bosons and the s-channel process mediated by the Higgs boson are suppressed.

\subsection{case-$H_5$}

\begin{figure}[!tbp]
\centering
\includegraphics[width=0.48\textwidth]{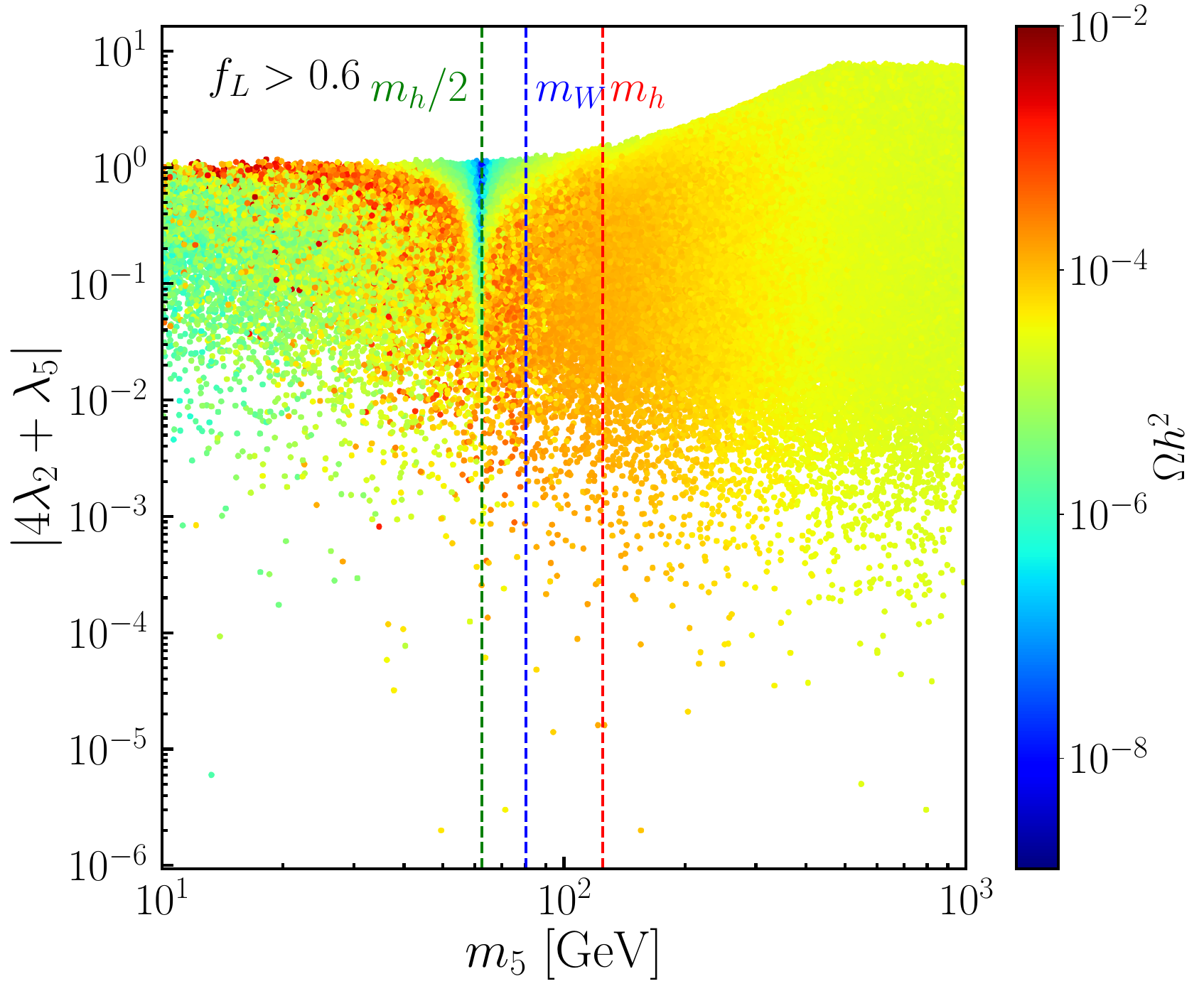}
\includegraphics[width=0.48\textwidth]{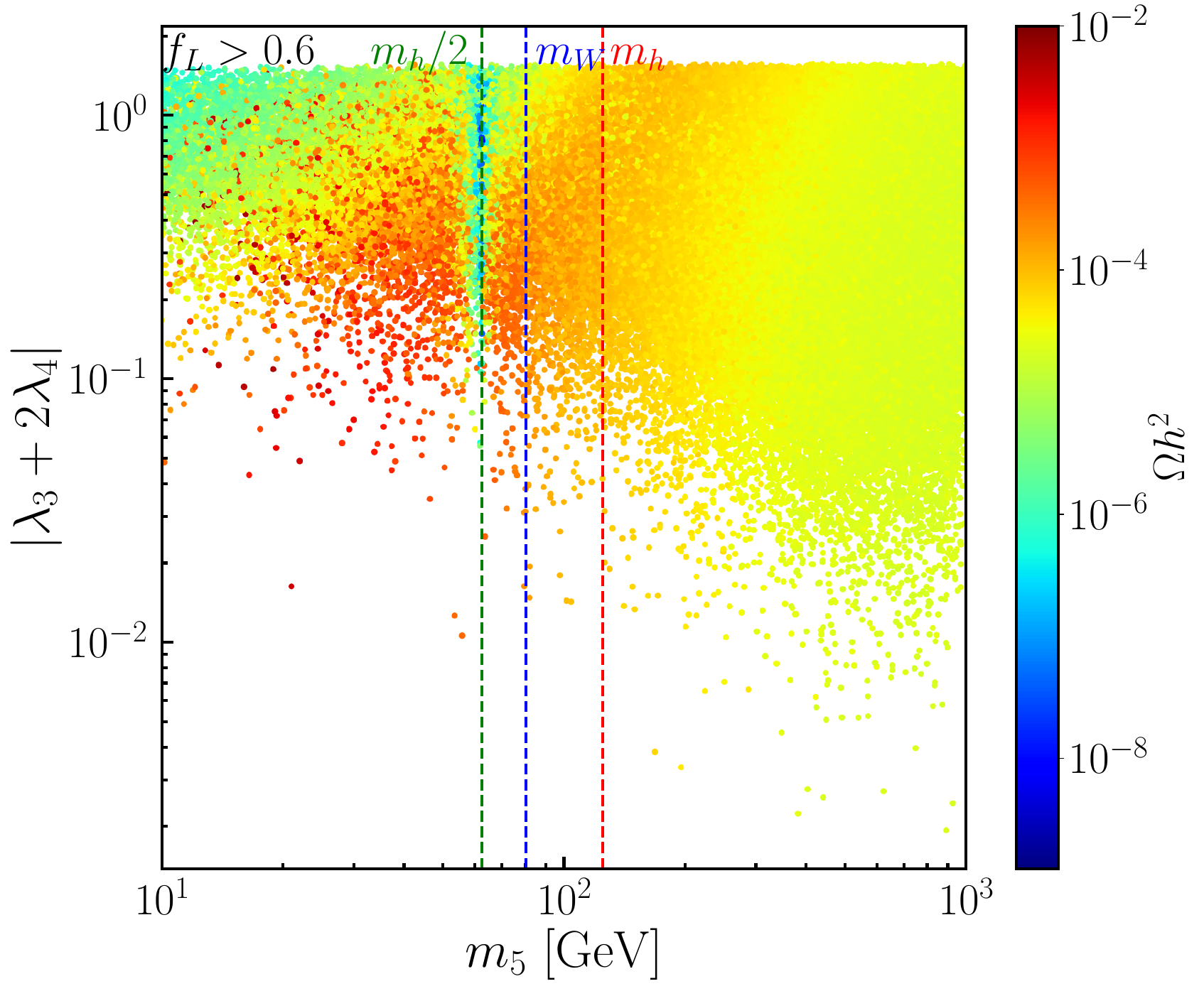}
\caption{The relic density of the DM in the $|4\lambda_2 + \lambda_5|$ vs. $m_5$ plane (left) and $|\lambda_3 + 2\lambda_4|$ vs. $m_5$ plane (right) for case-$H_5$ in the {\it L-Dominant} scenario.
}
\label{fig:case-H5-relic-LD-lam}
\end{figure}

For case-$H_5$, most situation is similar to that in case-$H$. However, the DM relic density in this case is always below the current measurement. This is due to the fact that {\it Proc-L} in this case contains more processes involving the other components in the fiveplet which are not suppressed by phase space in addition to the Higgs mediated $s$-channel processes and gauge boson mediated $s/t/u$-channel processes.
The dependence of the relic density in {\it L-Dominant} scenario for case-$H_5$ on the mass and couplings $4\lambda_2+\lambda_5$, $\lambda_3 + 2\lambda_4$ are shown in~\autoref{fig:case-H5-relic-LD-lam}.
Since the processes containing only fiveplet components are mixed with processes containing only DM and SM particles, the dependence are more involved.
Due to the fact that the annihilation of DM is dominated by the processes involving other fiveplet components in the low mass region ($m_5<m_h/2$) where the process involving only DM and SM particles are heavily suppressed, the relic density is mainly influenced by $g_{H_5H_5H_5H_5}\sim\lambda_3+2\lambda_4$ in the low mass region ($m_{5}<m_h/2$). Near the Higgs resonance region ($m_{5}\approx m_h/2$),
the Higgs mediated s-channel processes, especially
the reactions of DM annihilating into the SM particles
dominate over other reactions, so $g_{H_5H_5h}/v_\phi\sim4\lambda_2+\lambda_5$ is more relevant. In higher DM mass region, the processes involving gauge bosons are also involved which depend on the gauge couplings. Hence, the dependence on $\lambda_3+2\lambda_4$ and $4\lambda_2+\lambda_5$ becomes mild.

The {\it Mixed} scenario for case-$H_5$ is also similar to that for case-$H$. The processes in {\it Proc-H} become important when the mass splitting is small as shown in~\autoref{fig:delta-M-fl}. The relic density of the DM candidate is hence suppressed when the mass splitting is small. However, since the processes in {\it Proc-L} are not suppressed, especially for those containing only fiveplet components, the influence of the co-annihilation processes is not as significant as in case-$H$ as can be seen from~\autoref{fig:delta-M-fl} that the $f_L$ has a lower bound around $0.2$.

\section{Experimental constraints from DM searches}
\label{sec:constraints}
\subsection{Collider searches}

As discussed in previous section, the relic density of the DM candidate has different behavior in the {\it L-Dominant} and {\it Mixed} scenarios.
For {\it L-Dominant} scenario, the trilinear coupling between the DM and the Higgs ($g_{HHh}$, $g_{H_5H_5h}$) is important. Such coupling not only controls the relic density of the DM candidate, but also significantly affects the collider phenomenology as it is the portal coupling between the $Z_2$-odd sector and the SM.
On the other hand, the {\it Mixed} scenario in general indicates a small mass splitting $\Delta$ which is defined as the mass difference between the triplets and the DM candidate: $\Delta = m_3 - m_H$ for case-$H$ and $\Delta = m_3-m_5$ for case-$H_5$. However, when the mass splitting is sufficiently small, the collider phenomenology will also be altered when considering the production and decay of those slightly heavier scalars~\cite{ATLAS:2019lng}. In this work, $\Delta=2\,\rm GeV$~\cite{ATLAS:2019lng} is used as threshold for the collider analysis. Hence, in this section, $\Delta > 2\,\rm GeV$ and $\Delta <2\, \rm GeV$ will be separated when we examine the constraints from the collider searches on the $Z_2$ symmetric GM model.

\subsubsection*{Higgs invisible and exotic decays}

The Higgs-mediated $s$-channel processes contribute the most for the DM relic density which heavily depend on the trilinear coupling between the DM and the Higgs ($g_{HHh}$, $g_{H_5H_5h}$). However, the same coupling will significantly influence the decay of the Higgs when the mass of extra scalars, either the DM candidates or other scalars, is lighter than half of the Higgs mass. In the SM, the branching fraction of the invisible decay of the Higgs is about only $0.12\%$ which comes from $h\to ZZ^*\to \nu\bar{\nu}\nu\bar{\nu}$~\cite{ParticleDataGroup:2024cfk}. However, with the existence of the light DM candidate, the invisible decay of the Higgs can be enhanced significantly. Currently, the ATLAS and CMS collaborations have obtained the upper bound at 95\% CL on the invisible branching fraction assuming SM production as ${\rm BR}(h\to\text{invisible})<10.7\%$~\cite{ATLAS:2023tkt} and ${\rm BR}(h\to\text{invisble})<15\%$~\cite{CMS:2023sdw}, respectively. These measurements put strong constraints on the couplings between the DM candidate ($H$ for case-$H$, $H_5^0$ for case-$H_5$) and the SM Higgs.

On the other hand, the extra scalars that interact with the Higgs will also alter the corresponding decay patterns. However, if the mass difference between the extra scalar and the DM candidate is small, e.g. $\Delta<2\,{\rm GeV}$, the visible decay products from $h\to SS$ ($S=H,H_3,H_5$) are too soft to be detected by the detectors~\cite{ATLAS:2019lng}. It will still be considered as invisible channel. Otherwise, it will be treated as exotic decay of the Higgs providing different decay patterns from the existing SM channels. The analysis of ATLAS and CMS provide the upper bounds on the branching fraction of exotic decay of the Higgs as ${\rm BR}(h\to\text{undetected})<12\%$~\cite{ATLAS:2022vkf} and ${\rm BR}(h\to\text{undetected})<16\%$~\cite{CMS:2022dwd} at 95\% CL, respectively.

\begin{figure}[!tbp]
\centering
\includegraphics[width=0.48\textwidth]{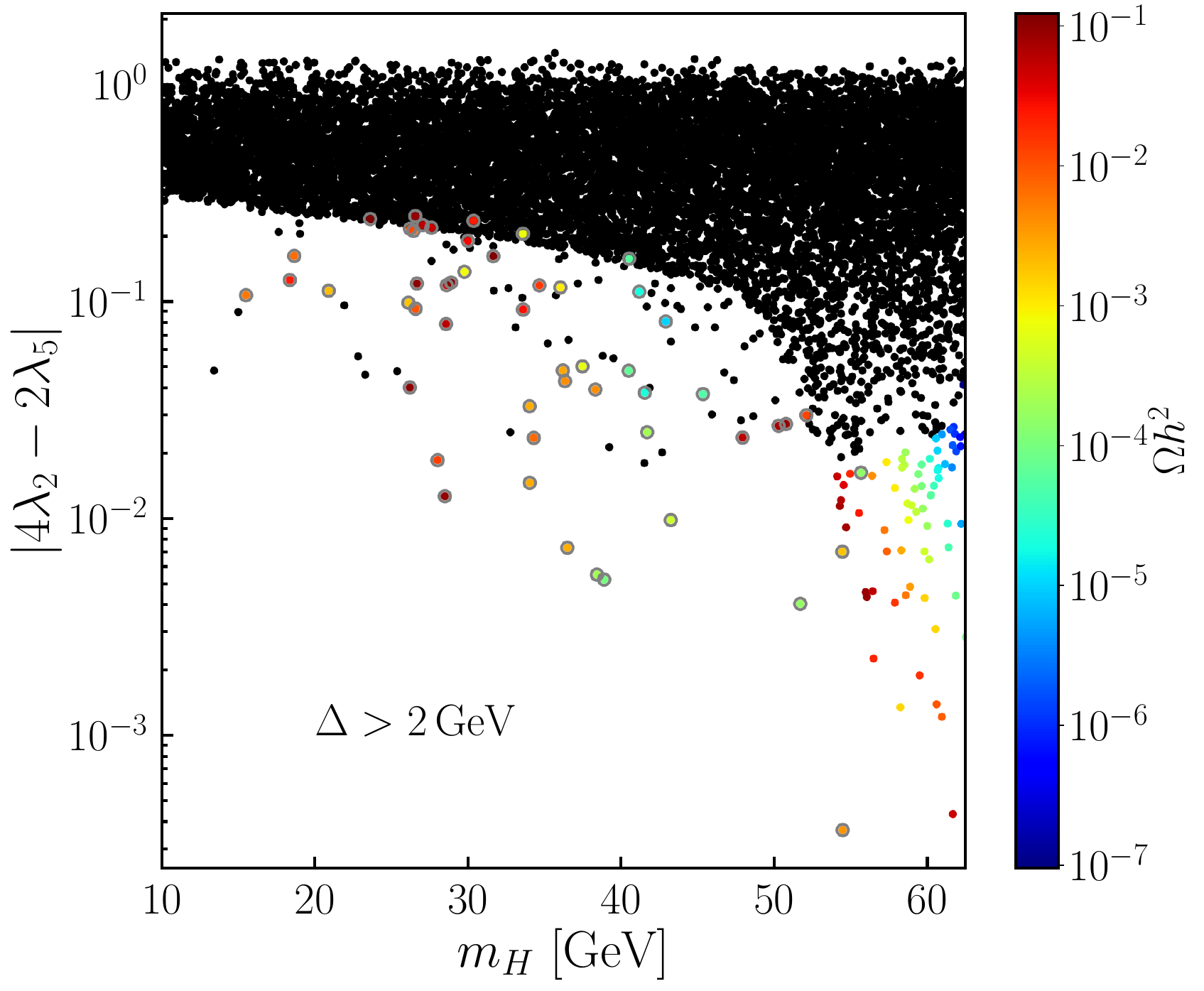}
\includegraphics[width=0.48\textwidth]{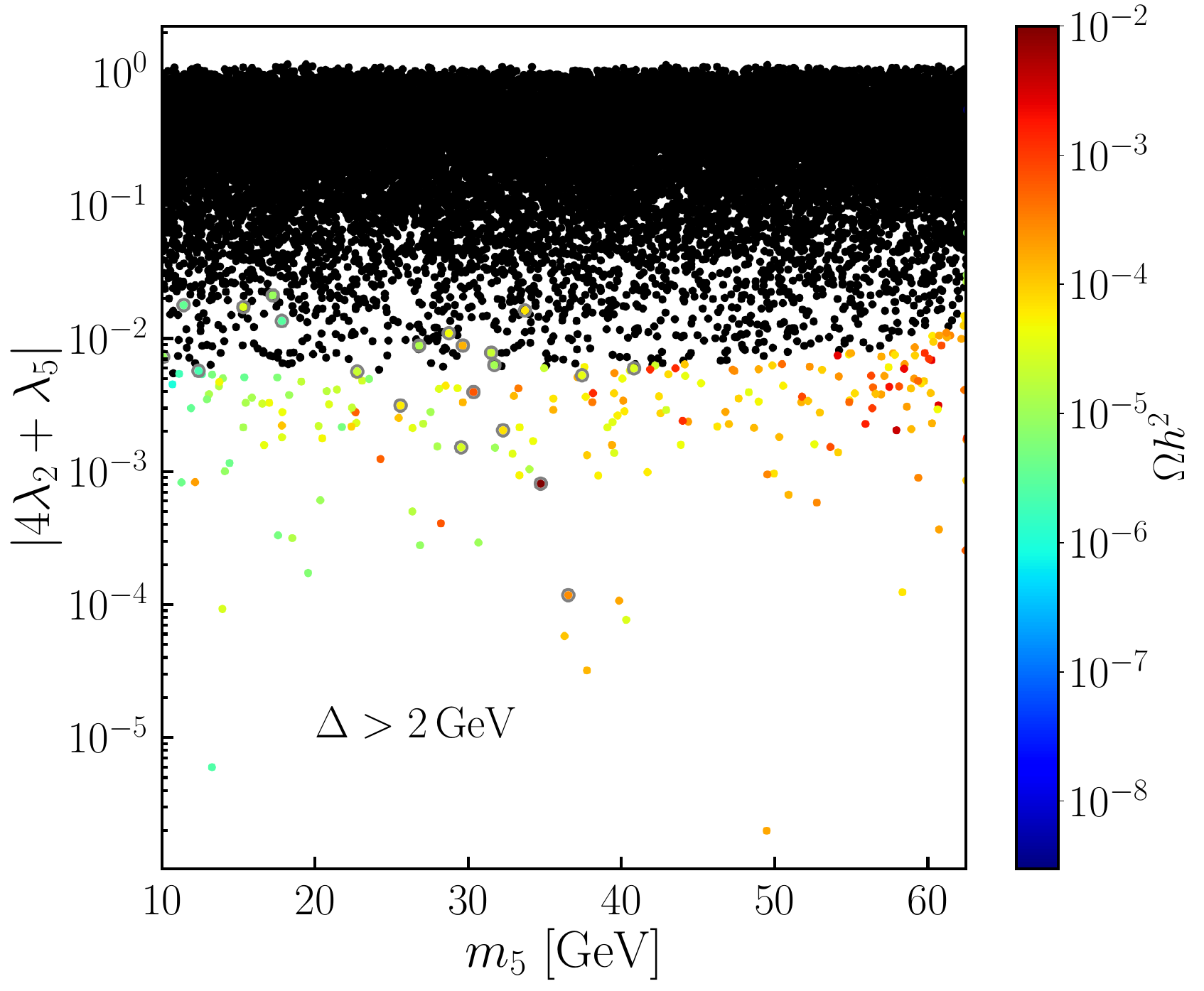}
\caption{\label{fig:Higgs-Delta-g}The constraints from the Higgs invisible and exotic decays on the trilinear coupling between the DM candidate and the Higgs for case-$H$ (left) and case-$H_5$ (right) for $\Delta > 2\,\rm GeV$.
The black points are excluded by the Higgs invisible decay. For the remaining points, The color of other points indicate the corresponding relic density. In addition, the gray circles represent the parameter points that are excluded by Higgs exotic decays but not by the Higgs invisible decays.
}
\end{figure}

In Fig.~\ref{fig:Higgs-Delta-g}, we show the constraints from the Higgs invisible and exotic decays on the trilinear coupling between the DM candidate and the Higgs for both case-$H$ (left) and case-$H_5$ (right) when $\Delta>2$ GeV. The black points represent those excluded by Higgs invisible decay measurements. While the gray circles represent those excluded by Higgs exotic decay measurement but not by Higgs invisible decay.
We find that the constraint from Higgs invisible decay has minor dependence on the DM mass and leads to $|g_{HHh}/v_\phi|=|4\lambda_2-2\lambda_5|\lesssim 2\times 10^{-2}$ for case-$H$ and $|g_{H_5H_5h}/v_\phi|=|4\lambda_2+\lambda_5|\lesssim 6\times 10^{-3}$. The constraints for case-$H_5$ are stronger as there are extra contributions to invisible decay from the other components of the fiveplet. The constraint from Higgs exotic decay will be more involved. The main contribution comes from $h\to H_3H_3$ which depends on $m_3$ and $g_{H_3H_3h}/v_\phi=4\lambda_2 - \lambda_5$.
It can exclude some parameter points with smaller $|g_{HHh}/v_\phi|$ in case-$H$ and $|g_{H_5H_5h}/v_\phi|$ in case-$H_5$ as indicated in~\autoref{fig:Higgs-Delta-g} by those gray circles.
In addition, the $H_3$ scalar of these points are relatively light. Hence, they will most probably be covered by the LEP constraints considered below.

When $\Delta<2$ GeV, we need to consider the contamination from the heavier scalars. The amended branching fraction of Higgs invisible decay is thus always be very large. Therefore, most points with $\Delta < 2\,\rm GeV$ and with mass below $m_h/2$ will be excluded by Higgs invisible decay measurements.

\subsubsection*{Mono-jet}

The mono-jet signature is a common signal for DM searches at the LHC~\cite{Godbole:2016mzr,ATLAS:2021kxv}. It is characterized by the final-state with at least one energetic jet and large $\slashed{E}_T$.
In the $Z_2$ symmetric GM model, the mono-jet signal is mainly generated by the Higgs boson mediating the production of a DM pair, accompanied by at least one energetic jet.
We recasted the current ATLAS mono-jet search by selecting events of DM pair production with one energetic recoil jet according to Ref.~\cite{ATLAS:2021kxv} using {\tt MadGraph}~\cite{Alwall:2014hca} to obtain the signal cross section. This cross section is then compared with the upper limit from ATLAS~\cite{ATLAS:2021kxv}. For $\Delta<2\,\rm GeV$, we also include the production of triplet pair with one energetic jet as signals.

For the $\Delta>2$ GeV, when $m_{DM}<m_h/2$, the DM pair can be produced resonantly with enhanced cross section. The mono-jet search puts upper limit on the trilinear coupling between the DM and the Higgs, $|g_{HHh}/v_\phi|=|4\lambda_2-2\lambda_5| \lesssim 3\times 10^{-2}$  for case-$H$ and $|g_{H_5H_5h}/v_\phi|=|4\lambda_2+\lambda_5| \lesssim 8\times 10^{-3}$ for case-$H_5$ with both nearly independent of $m_{DM}$. Due to similar reason as in Higgs invisible decay analysis, the constraints for case-$H_5$ is more stringent that that for case-$H$.
On the other hand, for higher $m_{DM}$, the constraints on the trilinear coupling between the DM and the Higgs ($g_{HHh}$, $g_{H_5H_5h}$) will be levitated as the cross section of the DM production through an off-shell Higgs is highly suppressed. The exclusion limit is even weaker than that from the theoretical constraints mentioned in~\autoref{sec2_2}.
For the $\Delta<2$ GeV, the mass of triplet $H_3$ is close to the DM mass, the production of triplet together with an energetic jet can also contribute to the mono-jet signature, then the restrictions will be slightly stronger.

\subsubsection*{Forward-backward dijet}

Forward-backward jet pairs production with large $\slashed{E}_T$ can also be used for the search of VBF production of DM pair. The DM pair will escape the detector and leave two energetic nearly back-to-back jets.
In this work, we calculated the cross-section for VBF production of DM pair in the $Z_2$ symmetric GM model by requiring two leading jets with $p^j_T > 20$ GeV, $|\eta_j| < 5.0$, $\Delta R_{j_1 j_2} > 0.4$, $|\eta_{j_1}-\eta_{j_2}| > 2.5$, and $\eta_{j_1}\eta_{j_2} < 0$.
The cross-section is then compared with the 95\% upper limit given by ATLAS~\cite{ATLAS:2022yvh}. Note that, when the DM mass is larger than $m_h/2$, the Higgs mediated contribution is suppressed. We thus only consider the situation when DM mass is smaller than $m_h/2$.

When $\Delta > 2$ GeV, only DM pair production is included. We obtain the upper limit on the coupling between DM and Higgs as: $|g_{HHh}/v_\phi|=|4\lambda_2 - 2\lambda_5| < 1.5\times 10^{-2}$ for case-$H$, and $|g_{H_5H_5h}/v_\phi|=|4\lambda_2 + \lambda_5|<7.0\times 10^{-3}$ for case-$H_5$. When $\Delta < 2$ GeV, the productions of triplets are also included. The constraints become stronger for both case-$H$ and case-$H_5$.

\subsubsection*{LEP}

\begin{figure}[!tbp]
\centering
\includegraphics[width=0.48\textwidth]{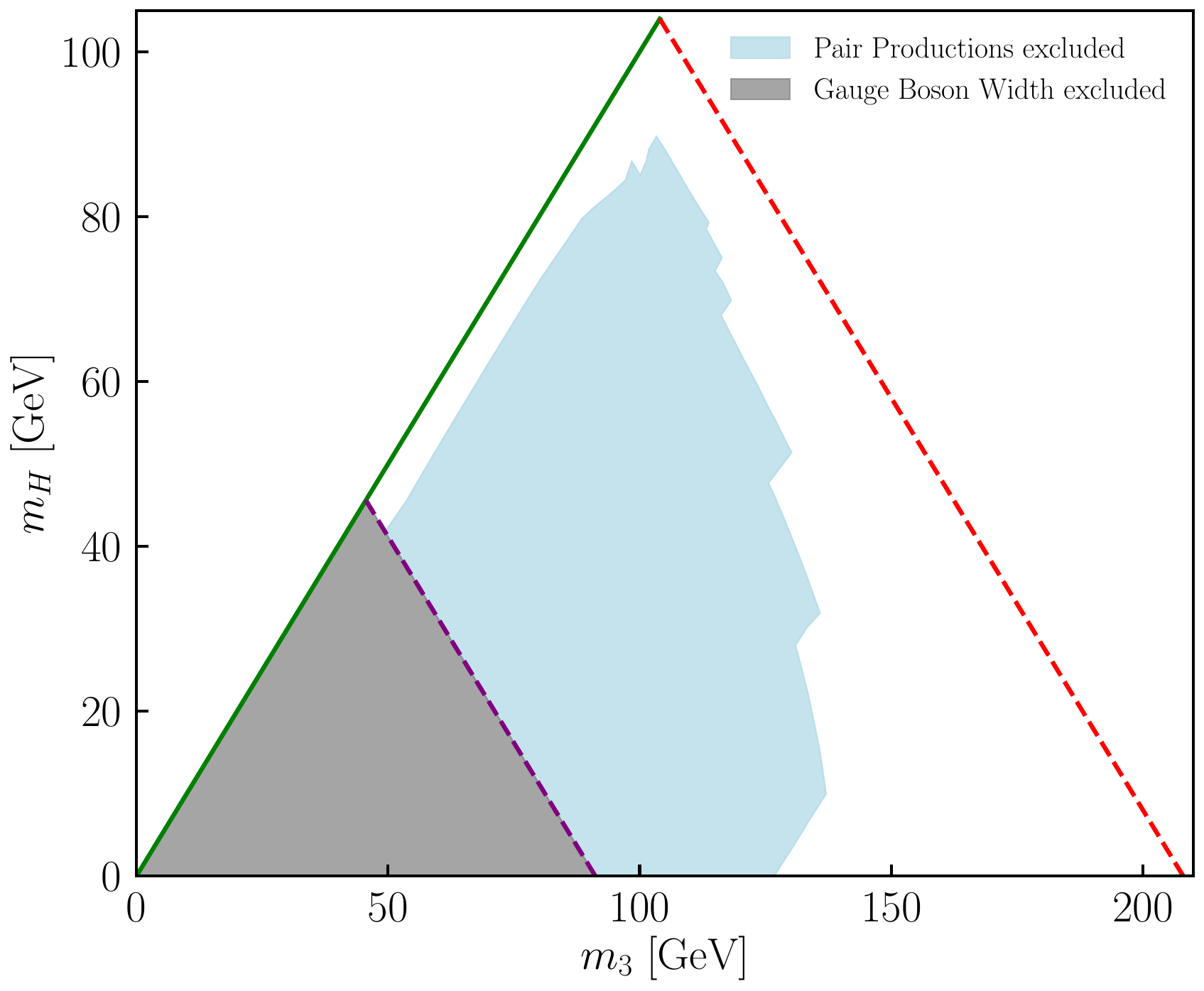}
\includegraphics[width=0.48\textwidth]{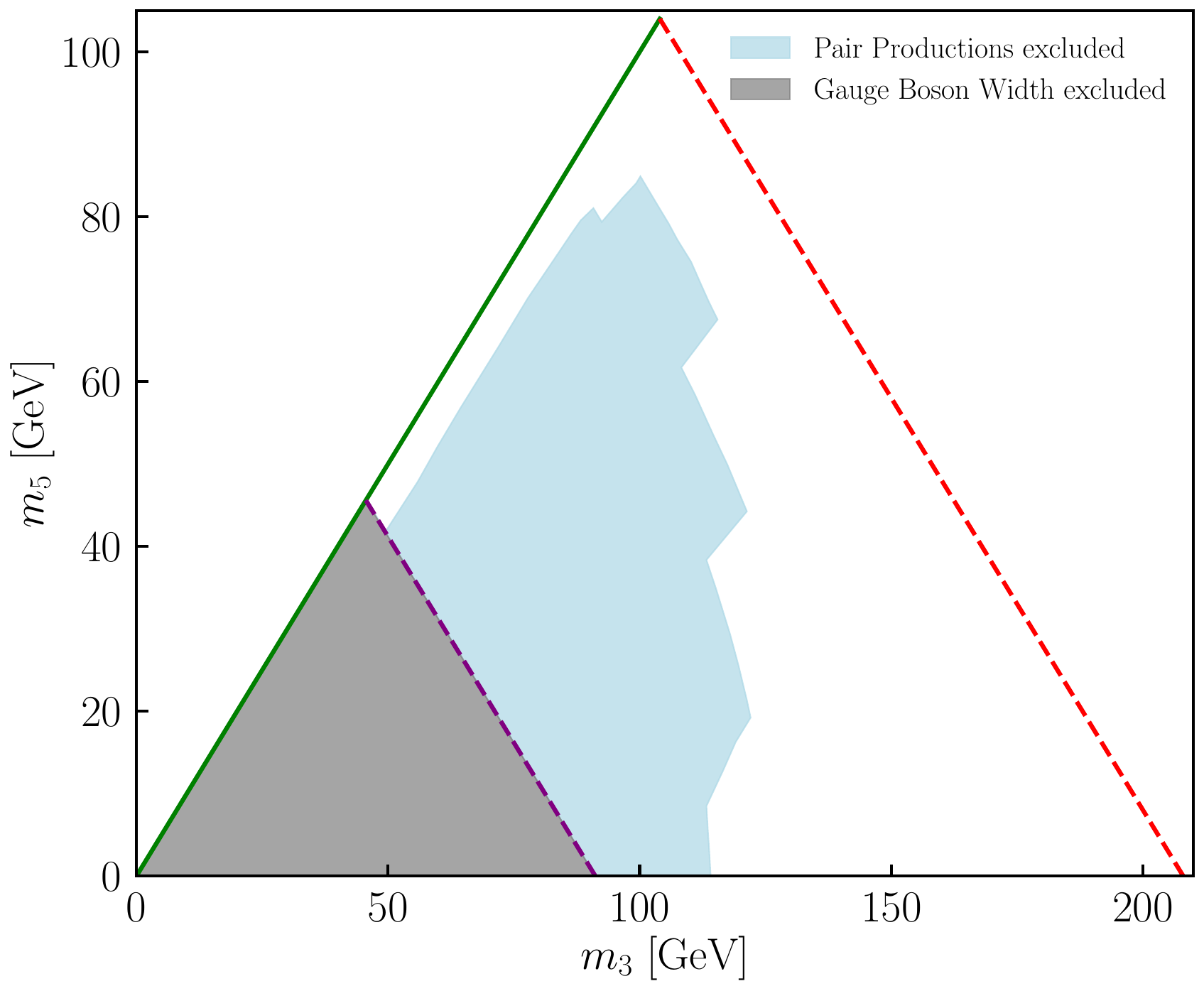}
\caption{The constraints from LEP measurement for case-$H$ (left) and case-$H_5$ (right). The lower left light gray region is excluded by the precise measurement of the $W/Z$ width. The blue region is excluded by the recast of the LEP II results considering the pair production of triplet. The red line is the kinematical limit of LEP II $m_{DM}+m_3 = 209\,\rm GeV$ and the green line is the condition $m_{DM} = m_{3}$.}
\label{fig:LEP}
\end{figure}

The LEP experiments precisely measured the properties of SM gauge bosons and searched for potential new particles around EW scale. These experimental results imposed stringent constraints on the masses and couplings of the new scalar particles ($H,H_3,H_5$) in the GM model. The well measured width of the W/Z boson~\cite{Lundstrom:2008ai,OPAL:2003wxm} imposed that the sum of the masses of the two scalars that couple to W/Z boson cannot exceed the mass of the corresponding gauge boson. For the $Z_2$ symmetric GM model, this imposed the following constraints
\begin{align}
    \label{equ:LEPI}
    \text{case-$H$}:&\quad m_{H} + m_{3} < m_{W,Z}, \qquad \text{case-$H_5$}:\quad m_5 + m_3 < m_{W,Z}.
\end{align}
For heavier scalar masses, the LEP II experiments~\cite{OPAL:2003nhx,OPAL:2003wxm,L3:1999onh,EspiritoSanto:2003by} also provided constraints on the production cross section of new particles which can be imposed on the $Z_2$ symmetric GM model. We recast the results from the DELPHI~\cite{EspiritoSanto:2003by} by considering $e^+ e^- \to H_3^+ H_3^-$ ($H_3^\pm \rightarrow l^\pm \nu S$) and $e^+ e^- \to H_3^0 S$ ($H_3^0 \to Z S$) where $S=H(H_5)$ for case-$H$ (case-$H_5$). The results are shown in~\autoref{fig:LEP} for both case-$H$ (left) and case-$H_5$ (right). The lower left gray region is excluded by~\autoref{equ:LEPI}. The light blue region is excluded by the searches of the pair production of the triplet.

\subsection{Direct detection}

\begin{figure}[!tbp]
    \centering
    \includegraphics[width=0.48\textwidth]{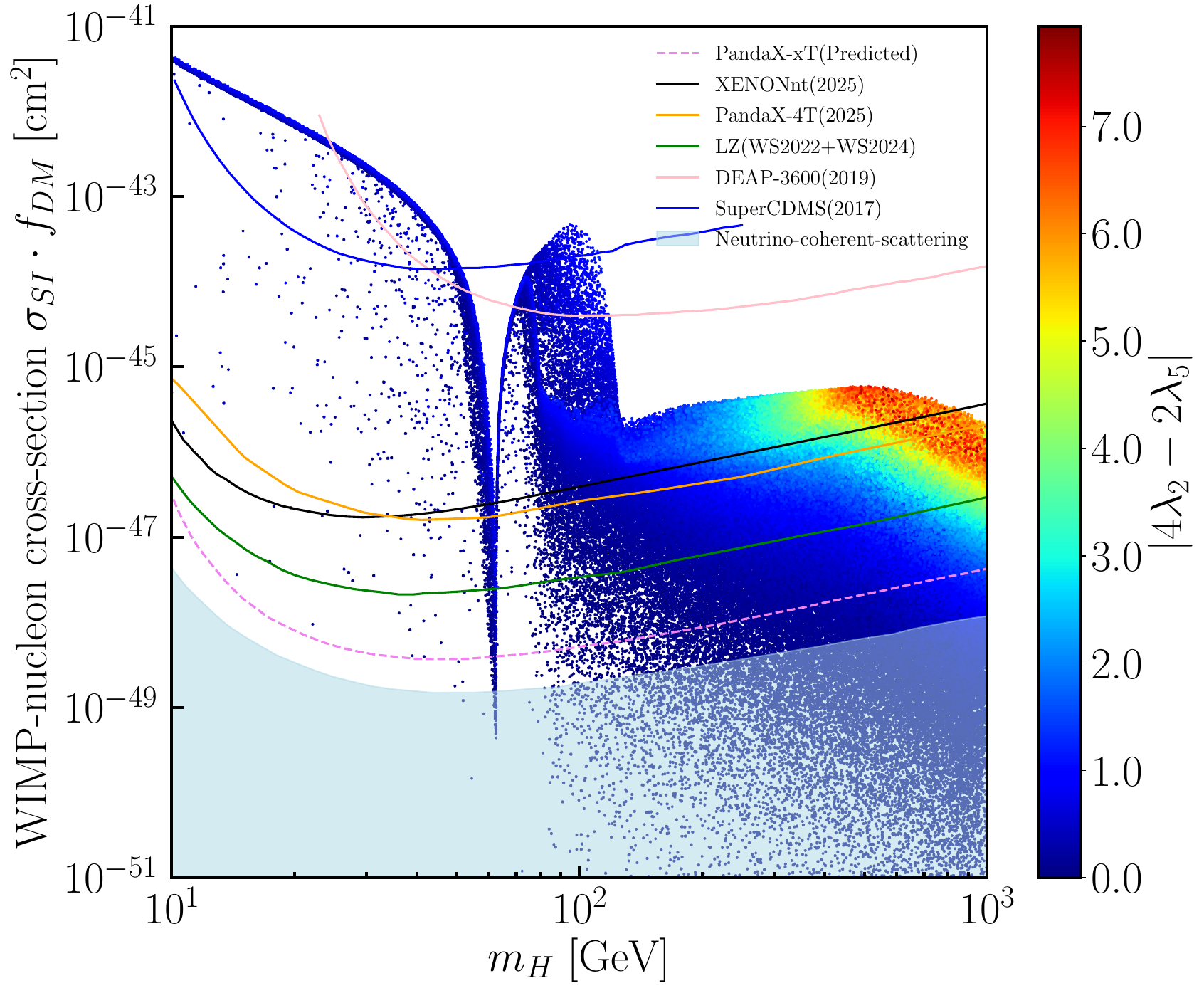}
    \includegraphics[width=0.48\textwidth]{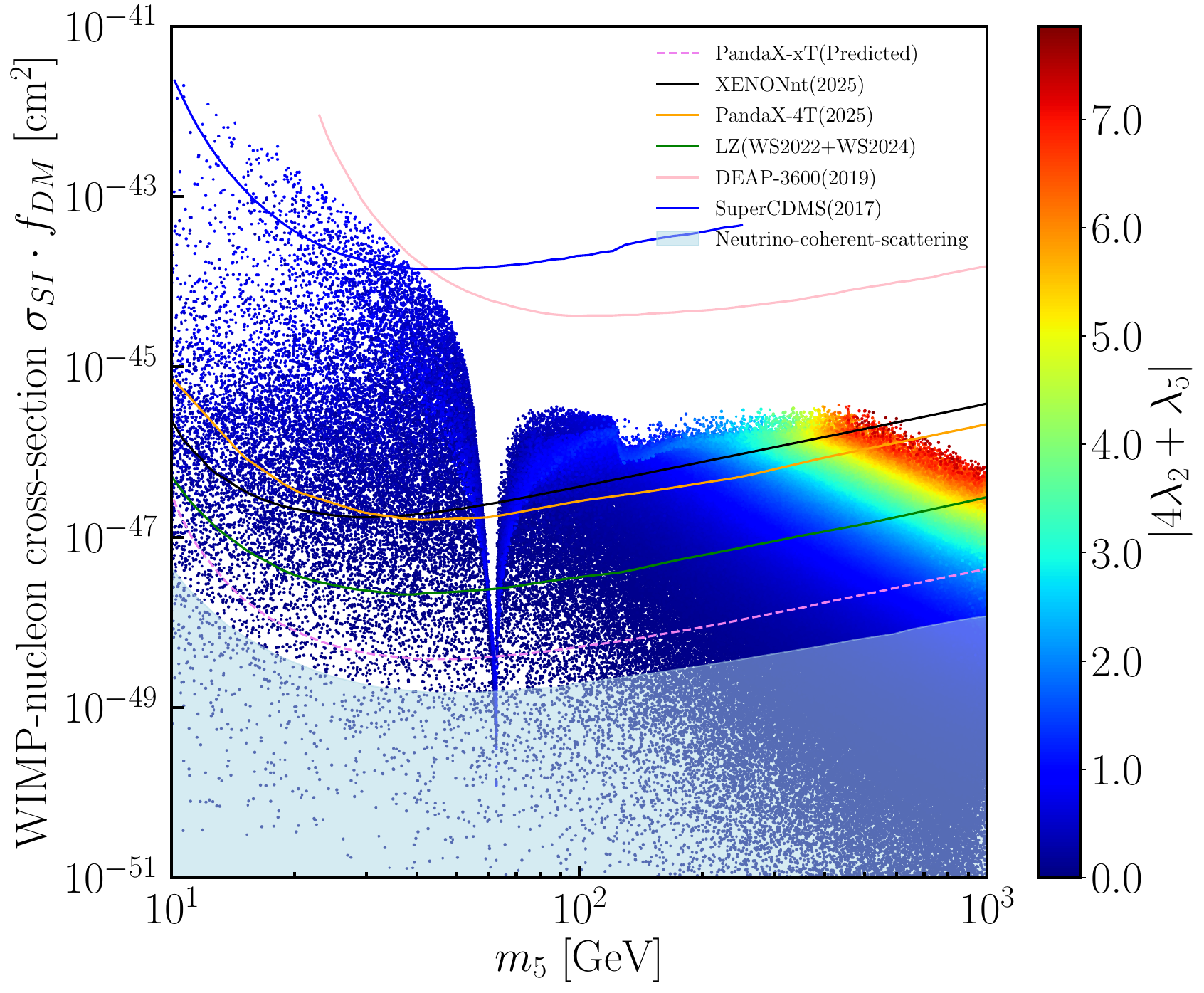}
\caption{The DM direct detection constraints from XENONnt~\cite{Aprile:2025egn}, DEAP-3600~\cite{DEAP:2019yzn}, SuperCDMS~\cite{SuperCDMS:2017mbc}, LZ~\cite{LZ:2024zvo}, PandaX-4T~\cite{PhysRevLett.134.011805} as well as the future prospect of PandaX-xT~\cite{PandaX:2024oxq} on the DM-nucleon cross section for the case-$H$ (left) and case-$H_5$ (right). The color of points indicate the correponding couplings $|4\lambda_2 - 2\lambda_5|$ for case-$H$ and $|4\lambda_2 + \lambda_5|$ for case-$H_5$. The parameter points that are over-abundant in case-$H$ are excluded. The parameter points that are under-abundant are rescaled according to the relic abundance $f_{DM} = \Omega_{th}/\Omega_{obs}$.
}
\label{fig:direct-detection}
\end{figure}

The DM direct detection imposes stringent constraints on the DM-nucleon interaction. In $Z_2$ symmetric GM model, the spin-independent DM-nucleon scattering cross section are shown in~\autoref{fig:direct-detection} for both case-$H$ and case-$H_5$ together with the exclusion limits from direct detection XENONnt~\cite{Aprile:2025egn}, DEAP-3600~\cite{DEAP:2019yzn}, SuperCDMS~\cite{SuperCDMS:2017mbc}, LZ~\cite{LZ:2024zvo}, PandaX-4T~\cite{PhysRevLett.134.011805} as well as the future prospect of PandaX-xT~\cite{PandaX:2024oxq}. The parameter points that are over-abundant in case-$H$ are excluded. Among the DM direct detection experiments, the latest LZ analysis combining the new 220 live-day exposure (WS2024)~\cite{LZ:2024zvo} with the 60 live-day exposure in the first result (WS2022)~\cite{LZ:2022lsv}, provides the most sensitive constraints to date. It is clear that the DM direct detection experiments can cover a large portion of the parameter space for both case-$H$ and case-$H_5$. Notice that for parameter points which are under-abundant, the DM-nucleon cross section is rescaled according to the relic abundance $f_{DM} = \Omega_{th}/\Omega_{obs}$.

 In $Z_2$ symmetric GM model, DM-nucleon interaction is mediated by the Higgs boson. The cross section is fully determined by the DM mass and the trilinear coupling between the DM candidate and the Higgs boson: $|g_{HHh}/v_\phi|=|4\lambda_2-2\lambda_5|$ for case-$H$ and $|g_{H_5H_5h}/v_\phi|=|4\lambda_2 + \lambda_5|$ for case-$H_5$. However, the event rate from the DM candidate in GM model should be rescaled according to its relic density which introduces more complicated dependence on the different scalar couplings and mass splitting. From~\autoref{fig:direct-detection}, it is clear that the DM direct detection impose very strong constraints on the parameter space of both case-$H$ and case-$H_5$ throughout the whole mass range.

\subsection{Indirect detection}

\begin{figure}[!btp]
\centering
\includegraphics[width=0.48\textwidth]{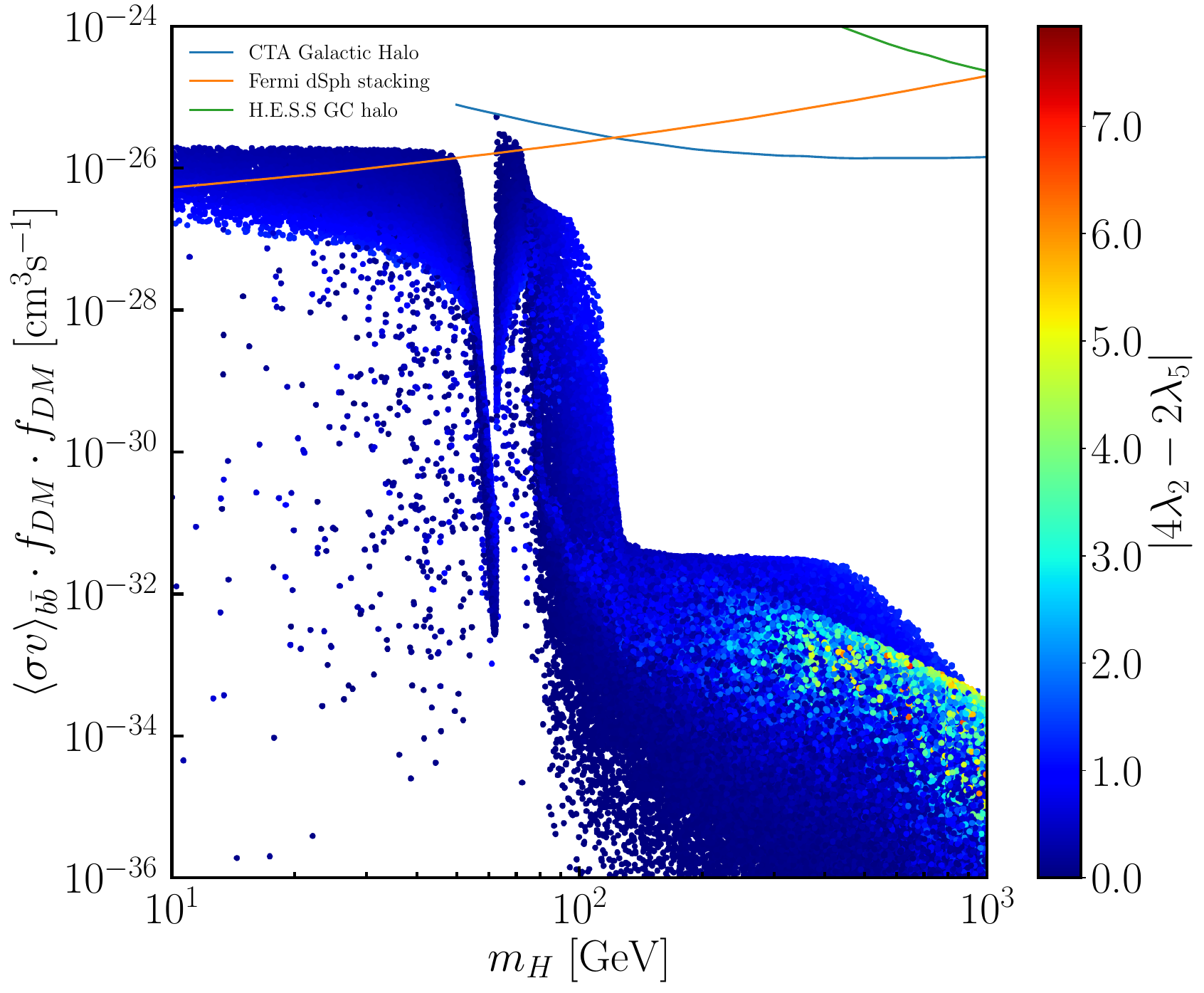}
\includegraphics[width=0.48\textwidth]{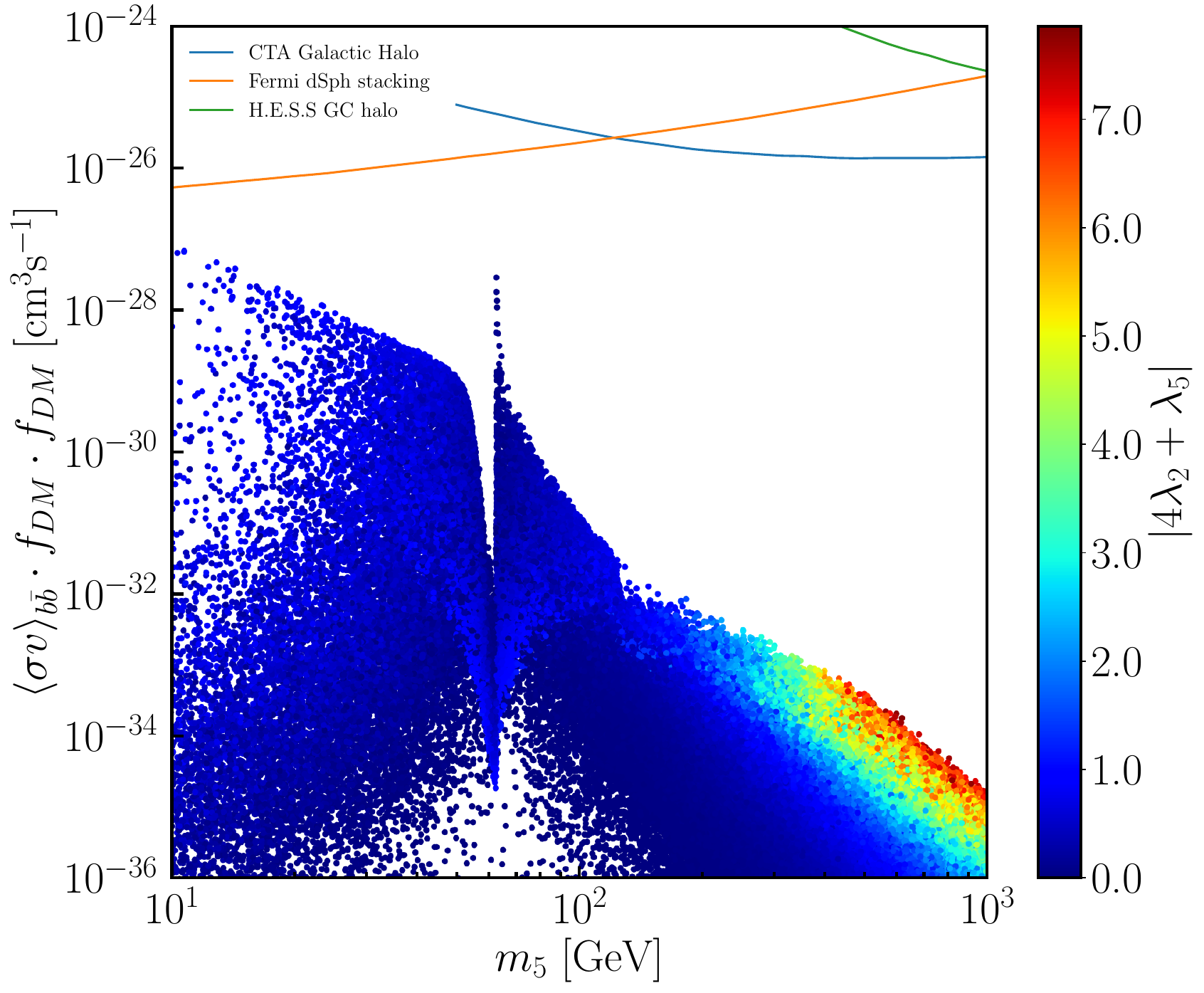}\\
\includegraphics[width=0.48\textwidth]{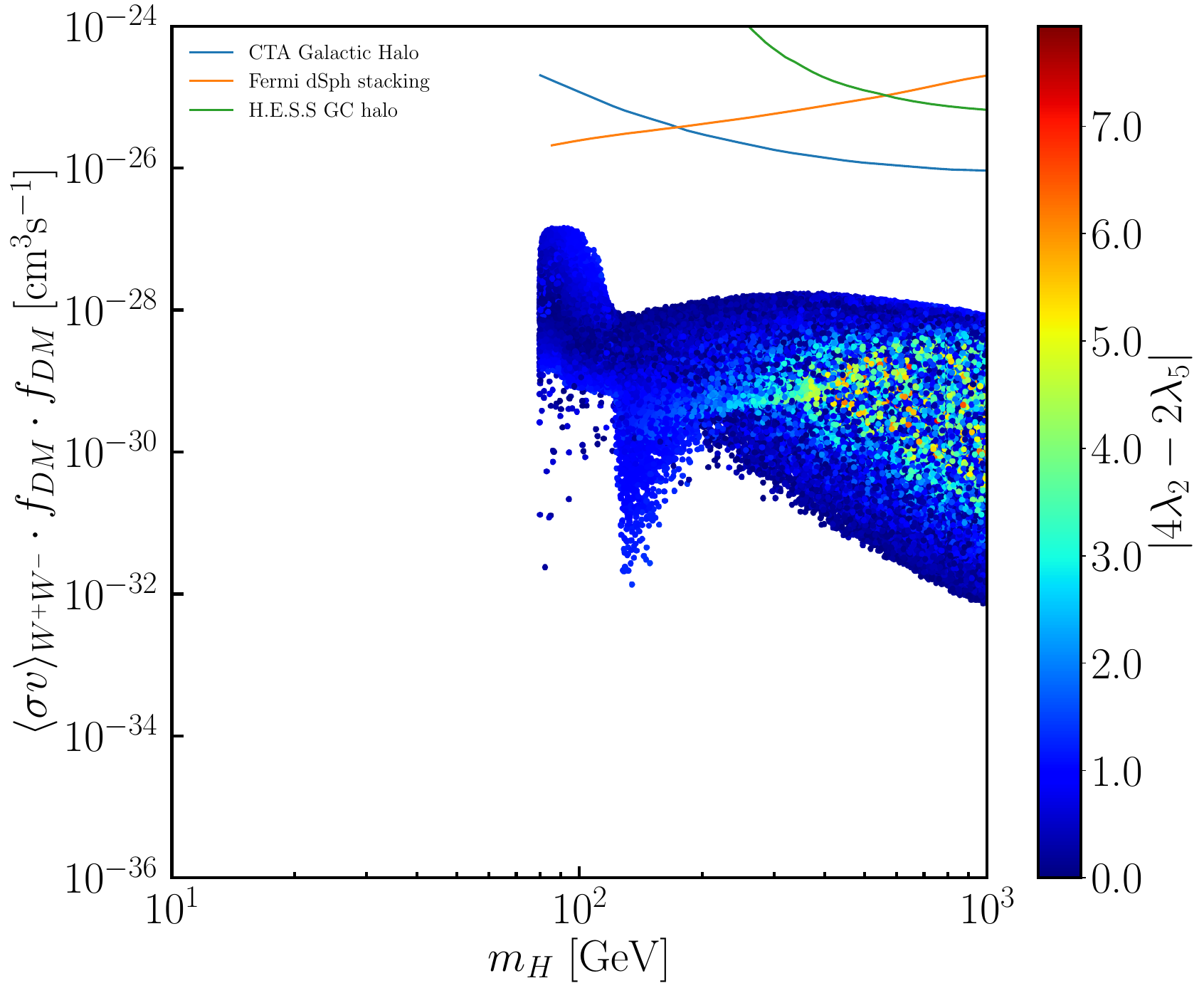}
\includegraphics[width=0.48\textwidth]{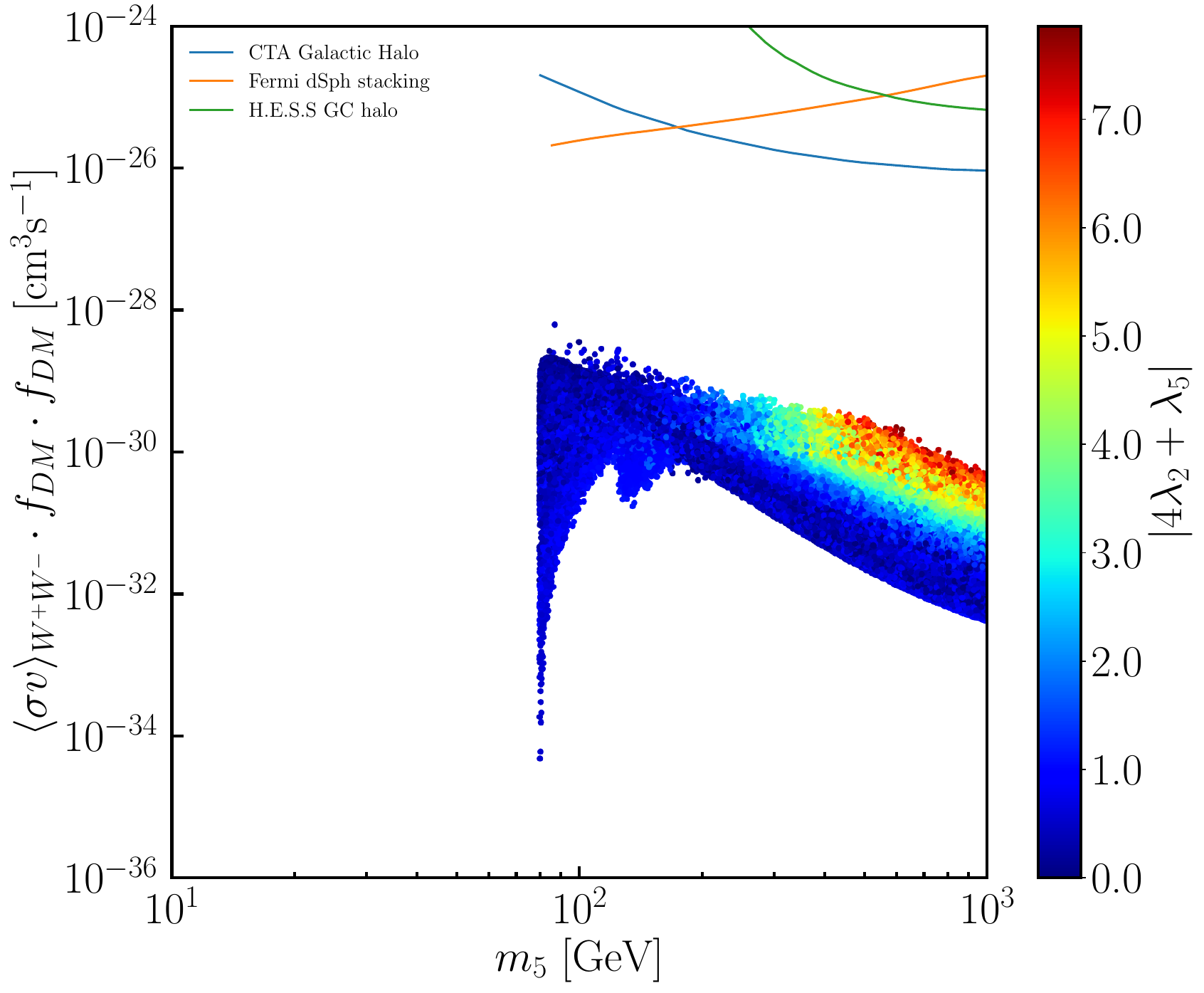}
\caption{Constraints on the DM annihilation cross-section for the $b\bar{b}$ (upper panels) and $W^+W^-$ (lower panels) channel for both case-$H$ (left panels) and case-$H_5$ (right panels). The upper limits on the annihilation cross-section from H.E.S.S.~\cite{HESS:2016mib} (green), Fermi-LAT~\cite{Fermi-LAT:2015att} (orange), and the sensitivity of the future CTA experiment~\cite{CTAConsortium:2017dvg} (blue) are also indicated.
}
\label{fig:ID-bb-ww}
\end{figure}

The DM indirect detection typically involves the annihilation of DM particles into SM particle pairs (bosons, quarks, or leptons), which subsequently produce observable secondary particles, such as gamma rays. These signals are particularly sensitive to the DM mass and its coupling strength with SM particles. Experimental data from telescopes such as Fermi-LAT~\cite{Fermi-LAT:2015att}
, H.E.S.S.~\cite{HESS:2016mib}
and CTA~\cite{CTAConsortium:2017dvg}, provide strong constraints on the DM annihilation cross-section.

The annihilation cross section of the DM into $b\bar{b}$ and $W^+W^-$ rescaled according to DM relic density in both case-$H$ and case-$H_5$ are shown in~\autoref{fig:ID-bb-ww} together with the current constraints from Fermi-LAT~\cite{Fermi-LAT:2015att}, H.E.S.S.~\cite{HESS:2016mib} and CTA~\cite{CTAConsortium:2017dvg} for each annihilation channel. Similar to DM direct detection, the annihilation processes are always mediated by the Higgs which strongly depend on $|g_{HHh}/v_\phi|=|4\lambda_2-2\lambda_5|$ for case-$H$ and $|g_{H_5H_5h}/v_\phi|=|4\lambda_2 + \lambda_5|$ for case-$H_5$. From~\autoref{fig:ID-bb-ww}, we can see that only the $b\bar{b}$ channel has sufficient annihilation cross section that can be probed by the Fermi-LAT results in the low mass region.

\subsection{EWPT and the Gravitational Wave Signal}

The necessary conditions for first-order EWPT are given in~\autoref{equ:SFOEWPT_Conditions}. In general, it requires larger quartic scalar couplings. Based upon these conditions, we implement the $Z_2$ symmetric GM model in {\tt CosmoTransitions}~\cite{Wainwright:2011kj}.
For strong first-order EWPT, the gravitational wave can be generated during the vacuum bubble expansion, which has been discussed in~\autoref{sec:EWPT}.
The Signal-to-Noise Ratio (SNR) is a key metric for evaluating the ability of gravitational wave detectors to identify signals. It can be calculated for a particular experiment and an expected GW spectrum through~\cite{Caprini:2015zlo}
\begin{equation}
\mathrm{SNR}=\sqrt{\mathcal{T} \int_{f_{\min }}^{f_{\max }} d f\left[\frac{h^{2} \Omega_{\mathrm{GW}}(f)}{h^{2} \Omega_{\mathrm{Sens}}(f)}\right]^2},
\end{equation}
where $f_{min}$, $f_{max}$ denote the minimal and maximal frequencies accessible of the detector, respectively. $h^{2} \Omega_{\mathrm{Sens}}(f)$ denotes the sensitivity of a given configuration of the experiment, $\mathcal{T}$ is the duration of the mission and $h^{2} \Omega_{\mathrm{GW}}(f)$ represents the possible GW signal. In this work, we assume $\mathcal{T}$ is 3 years, based on a four-year mission duration with 75$\%$ data-taking up time for the LISA mission~\cite{Caprini:2019pxz}.
The higher SNR indicates a stronger and more detectable signal. Therefore, we will use 10 as the threshold for GW signal detection.

\begin{figure}[!tbp]
    \centering
    \includegraphics[width=0.48\textwidth]{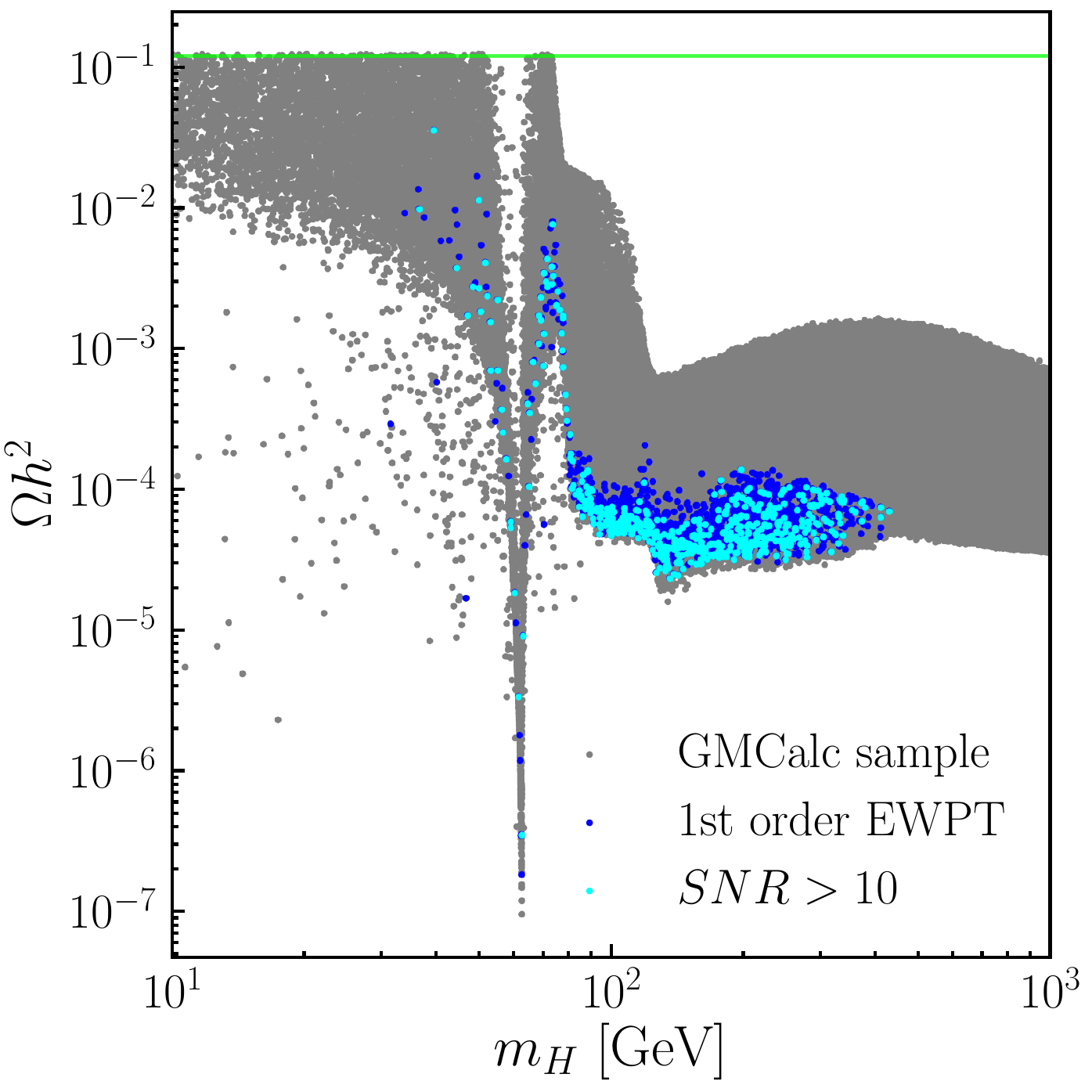}
    \includegraphics[width=0.48\textwidth]{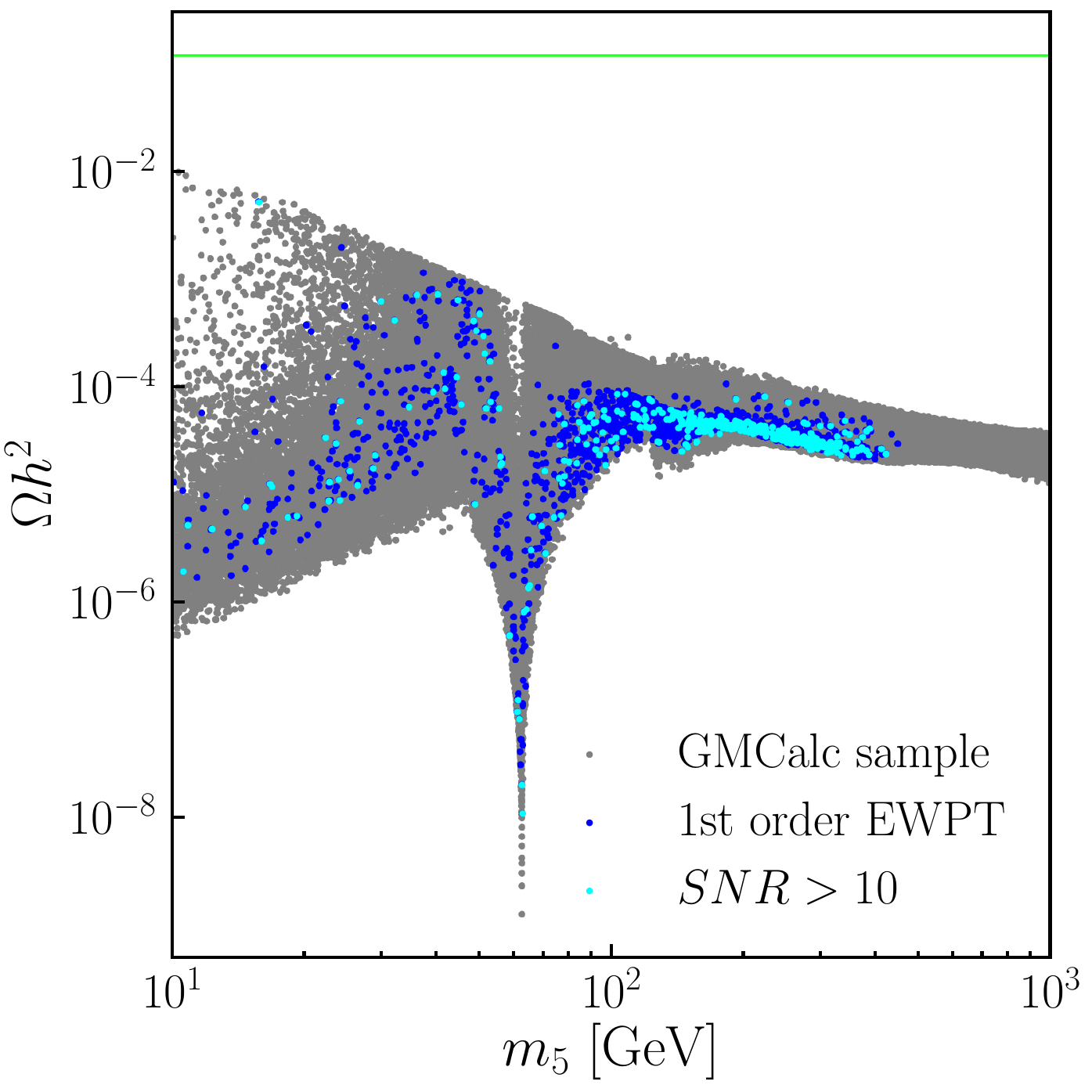}
    \caption{Relic density $\Omega h^2$ as a function of the DM mass $m_H$ (left) and $m_{H_5^0}$ (right). All parameter points that cannot trigger first-order EWPT are shown in gray. The blue points indicate those can trigger first-order EWPT and cyan points indicate those with an SNR greater than 10. The green horizontal band is the same as in Fig.~\ref{fig:Omega-M}.}
\label{fig:EWPT}
\end{figure}

\begin{figure}[!tbp]
    \centering
    \includegraphics[width=0.48\textwidth]{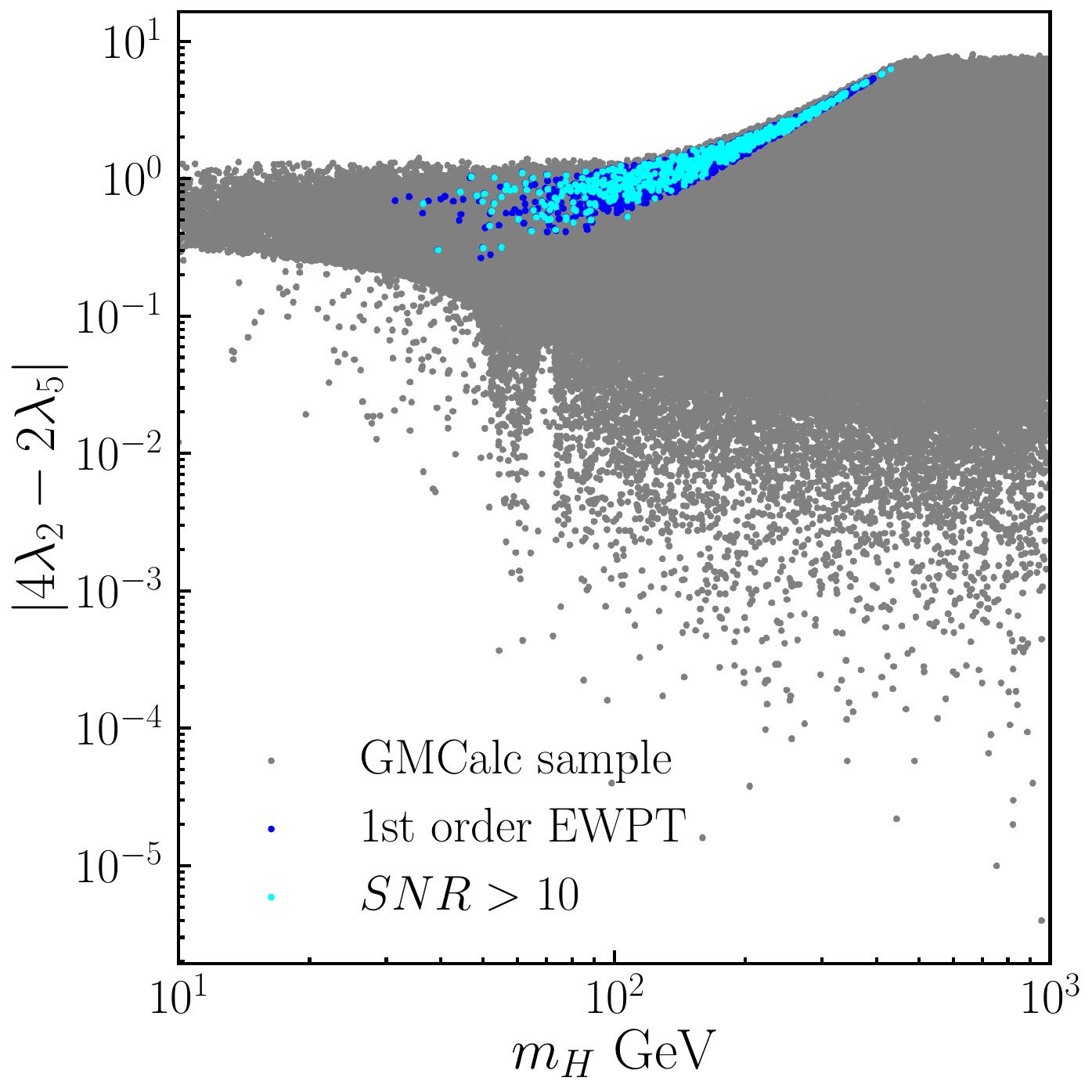}
    \includegraphics[width=0.48\textwidth]{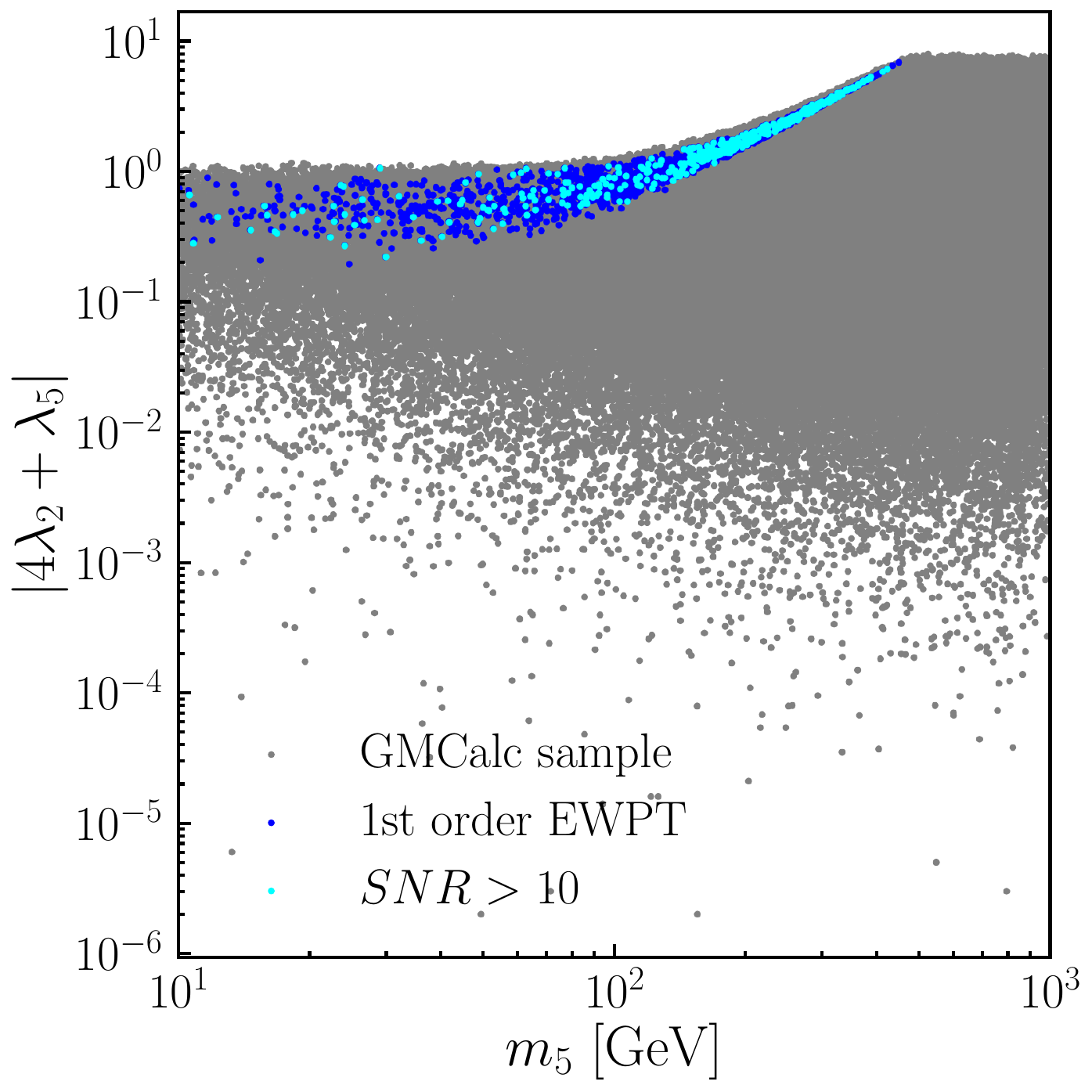}
    \caption{The trilinear coupling between the Higgs boson and DM candidates as a function of the DM mass $m_H$ (left) and $m_{H_5^0}$ (right). The colors of points are similar to~\autoref{fig:EWPT}.}
\label{fig:EWPT-lam}
\end{figure}

\autoref{fig:EWPT} displays the parameter points that yield a first-order phase transition for both case-\(H\) and case-\(H_5\). Only the blue points trigger a first-order EWPT, while the cyan points indicate those that can produce a detectable GW signal (SNR $>10$).
It is clear that the parameter points that can achieve first-order EWPT have relatively low relic density. This indicates that the DM has a relatively large annihilation cross section in this case. Furthermore, according to~\autoref{fig:EWPT-lam}, it is evident that strong gravitational wave signals indeed tend to prefer the region with large couplings between the DM candidates and the Higgs boson. This makes the parameter space in this part more prone to being excluded by other detections with the exception of the parameter region around the Higgs resonant.

\subsection{Summary of the Constraints}

In this subsection, we present brief summaries of all the constraints we have discussed above, among which the direct detection of DM appears to be the most stringent constraints. Since the constraints from the LHC (Higgs invisible/exotic decay, mono-jet, forward-backward jets) are different for $\Delta>2$ GeV and $\Delta<2$ GeV, the summaries presented here will also be separated accordingly.

\begin{figure}[!tbp]
    \centering
    \includegraphics[width=0.48\textwidth]{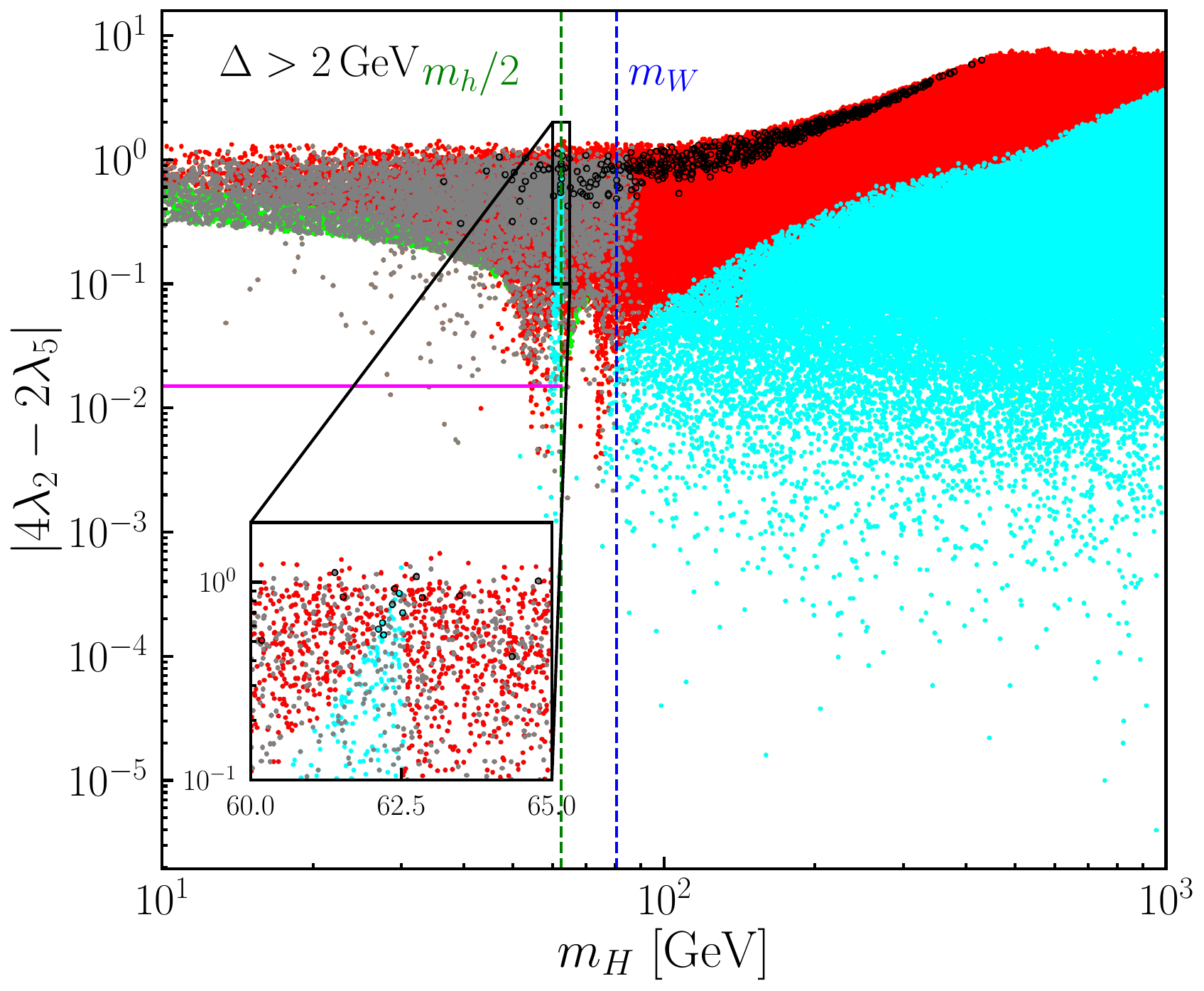}
    \includegraphics[width=0.48\textwidth]{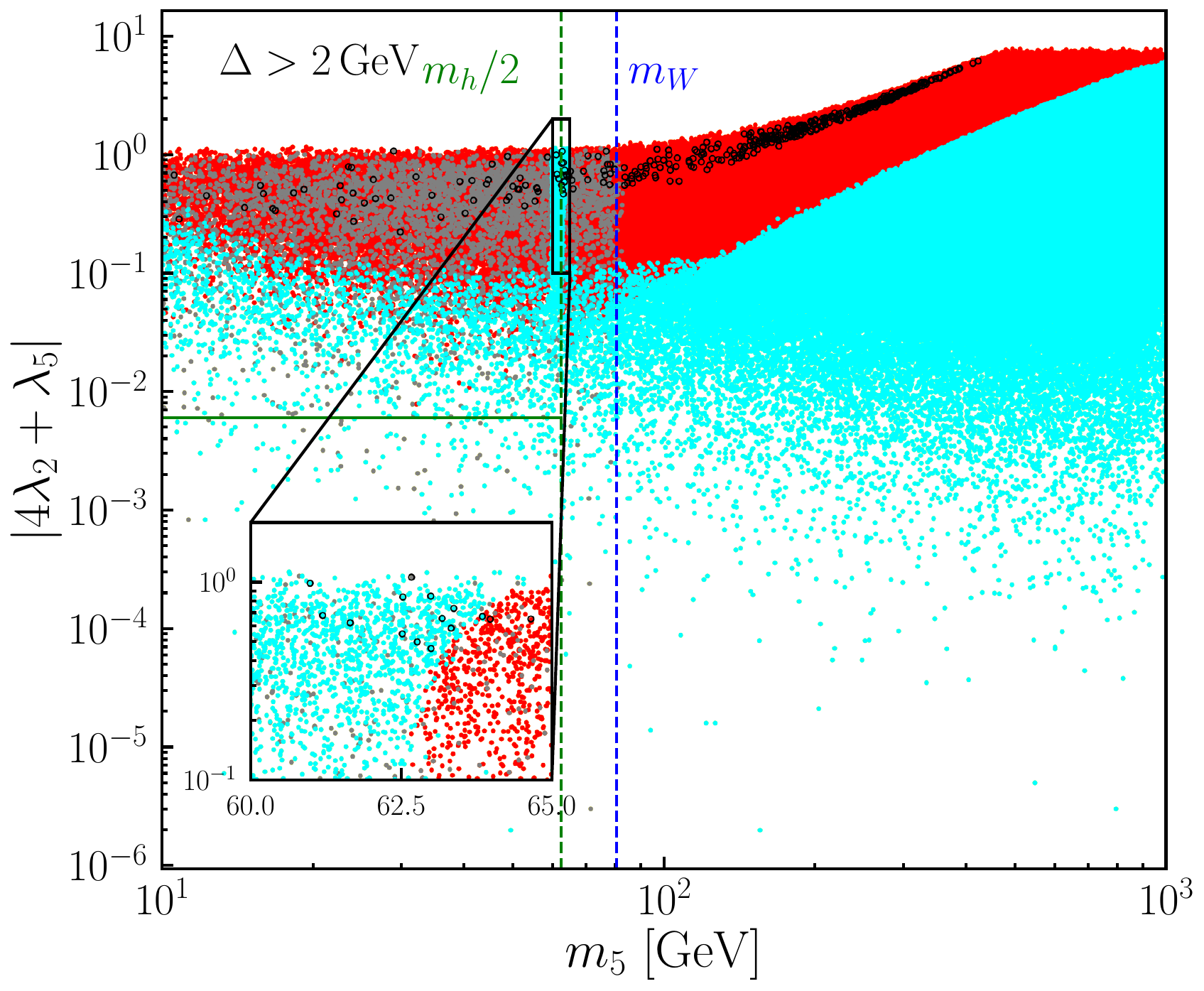}
    \caption{The constraints from the LHC, LEP, direct/indirect detection of DM and gravitational wave signals for both case-$H$ and case-$H_5$ with $\Delta>2$ GeV. The more details about the color code can be found in the main text. The small panels in both figures zoom out the Higgs resonant region.}
    \label{fig:full_constr_delta_gtr_2}
\end{figure}

The results for $\Delta>2$ GeV are shown in~\autoref{fig:full_constr_delta_gtr_2} for both case-$H$ and case-$H_5$. Note that for case-$H$, the parameter points that are over-abundance are not included in the plot. In this case, the constraints from the LHC are very predictive. For case-$H$, the measurement through VBF production of DM pair provides the strongest constraint as $|g_{HHh}/v_\phi|=|4\lambda_2-2\lambda_5|<1.5\times 10^{-2}$ for $m_H < m_h/2$. For case-$H_5$, it is the Higgs invisible decay that provides the strongest constraint as $|g_{H_5H_5h}/v_\phi|=|4\lambda_2+\lambda_5|<6\times 10^{-3}$ for $m_5 < m_h/2$. These are shown as magenta (for case-$H$) and green (for case-$H_5$) horizontal lines in~\autoref{fig:full_constr_delta_gtr_2}. The LEP experiments excluded a large portion of the parameter space when $m_3\lesssim 130\,\rm GeV$ for case-$H$ and $m_3\lesssim 110\,\rm GeV$ for case-$H_5$. These are shown in~\autoref{fig:full_constr_delta_gtr_2} with gray points. In addition to the LEP constraints, the red and blue points are further excluded by the DM direct and indirect detection, respectively. The parameter points that will be covered by both DM direct and indirect detection are shown in light green. The cyan points are those can escape all above constraints except the LHC constraints. Further, points with black edge are those with GW $\text{SNR}>10$. We find that the collider searches mostly concentrated on the low mass region, while the DM direct detection can provide strong constraints across a larger mass range. The points with strong GW signals usually have strong scalar couplings and hence will be mostly covered by the DM direct detection and/or collider searches. The only exception is the Higgs resonant region which is shown explicitly with zoom-in small panels in~\autoref{fig:full_constr_delta_gtr_2}, where all those constraints have no low sensitivity. However, it is still possible to probe part of this region by future GW detections.

\begin{figure}[!tbp]
    \centering
    \includegraphics[width=0.48\textwidth]{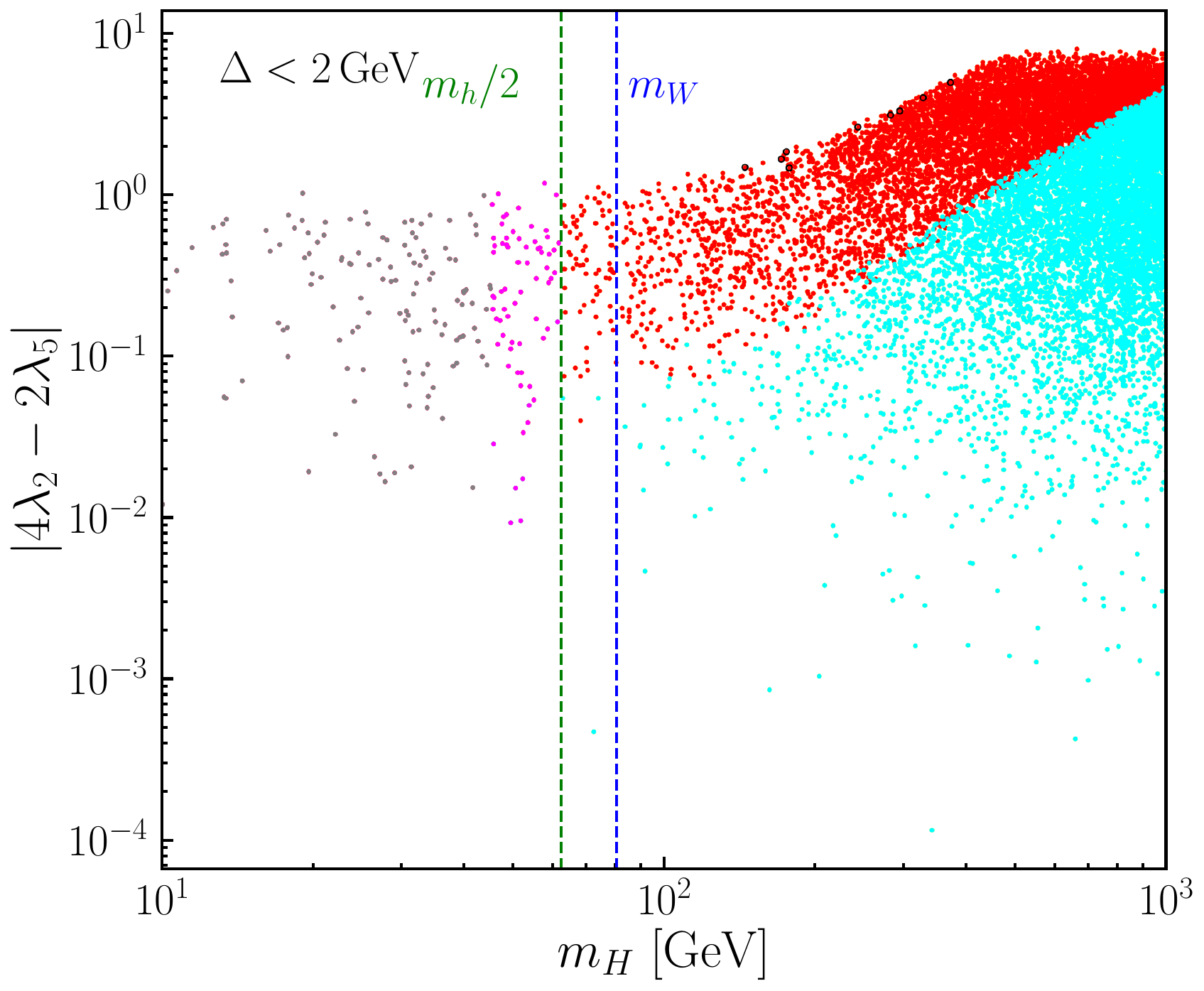}
    \includegraphics[width=0.48\textwidth]{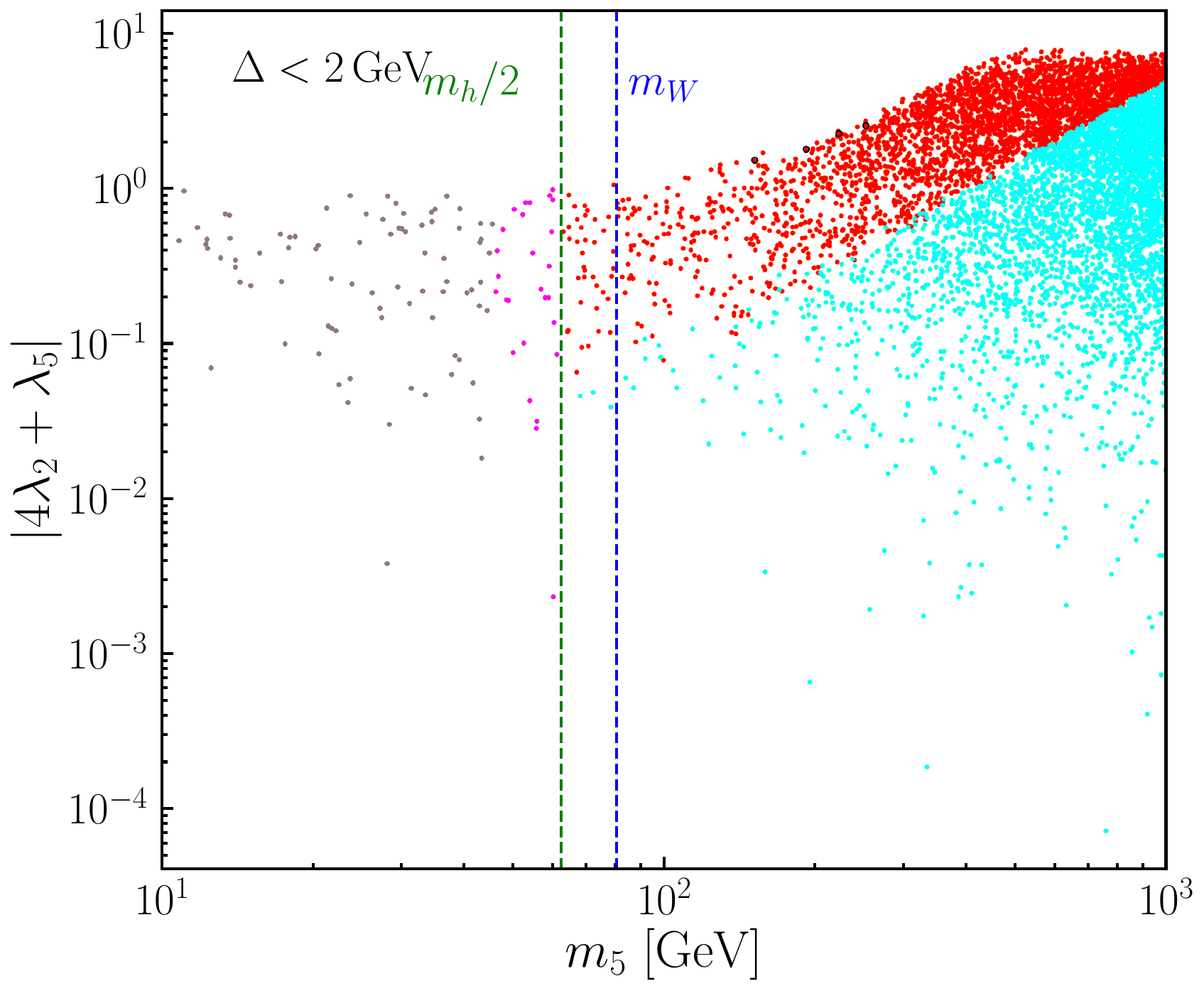}
    \caption{The constraints from the LHC, LEP, direct detection of DM and gravitational wave signals for both case-$H$ and case-$H_5$ with $\Delta<2$ GeV. The more details about the color code can be found in the main text.}
    \label{fig:full_constr_delta_lt_2}
\end{figure}

The results for $\Delta<2$ GeV are shown in~\autoref{fig:full_constr_delta_lt_2} for both case-$H$ and case-$H_5$. In this case, the constraints from LEP represented by gray points become weak, as there is no sensitivity from the pair productions of the triplets if $\Delta<2$ GeV. In addition to the LEP constraints, the LHC constraints, which are more involved than that of $\Delta>2$ GeV, are represented by the magenta points. Upon this, extra points that can be probed by DM direct detection are indicated as red. For $\Delta<2$ GeV, there is no sensitivity from the DM indirect detection. Further, the points with GW $\text{SNR}>10$ are shown with black edge. We find that in the parameter space where $\Delta < 2$ GeV, for both case-$H$ and case-$H_5$, due to the strong constraints imposed by collider experiments, the viable parameter space is mainly concentrated in the region with a relatively large DM mass.

\section{Conclusion}
\label{sec:conclusion}
In this work, we examine the DM properties and EWPT in $Z_2$ symmetric GM model. In the $Z_2$ symmetric GM model, when the vacuum also preserves the $Z_2$ symmetry, the scalar components from the $SU(2)_L$ triplet, either $H$ or $H_5^0$ can be DM candidates. In case-$H$, when $m_H<m_W$, the annihilation processes of the DM controlled by gauge couplings are suppressed by the phase space. Then, the observed relic density can be achieved through the annihilation processes mediated by the Higgs. Such processes depend on the coupling between the DM and the Higgs $g_{HHh}$. On the other hand, when $m_H$ is large, $HH\to WW$ dominates, the relic density of the DM is heavily suppressed. For case-$H_5$, with other components from the custodial fiveplet, the relic density of the DM is always suppressed and well below the observed value.

For these extra scalars, the collider experiments provide important probes to explore the parameter space. In this work, we extensively examine the constraint from LEP, including that from the gauge boson width measurements and that from the scalar production measurements, the constraint from the LHC, including the Higgs invisible decay, the mono-jet measurement and the VBF production of DM pair. These measurements are more sensitive to the low mass region, where the mass of the DM is smaller than half of the Higgs mass. For heavier mass, there is no sensitivity from LEP and LHC. It is mainly due to that the mass is beyond the reach of the LEP, and beyond the Higgs decay threshold. The processes of other measurements at the LHC are mediated by the Higgs. For heavier DM masses, the Higgs becomes off-shell, leading to a significant suppression of the production cross-section.

Among all the experiments, the DM direct detection provide the strongest constraints on the parameter space. It provides an upper limits on the couplings between DM and the Higgs ($g_{HHh}$, $g_{H_5H_5h}$) for a large range of the DM mass. For case-$H$, only a tiny region around Higgs resonant ($m_H\sim m_h/2$) and $m_H\gtrsim 100\,\rm GeV$ with relatively small couplings $g_{HHh}$ can escape the constraints of the DM direct detection. The situation is similar for case-$H_5$. However, due to the relic density in this case is heavily suppressed, parameter points of low mass $m_5\lesssim m_h/2$ but with relatively large couplings $g_{H_5H_5h}$ can also escape the direct detection constraints. The constraint from indirect detection is much weaker, as the annihilation of the DM is suppressed mainly due to the small Yukawa couplings for SM fermion channel, and the suppressed phase space or suppressed relic density for $WW$ channel.

We also explore the possibility of a first-order EWPT in the $Z_2$ symmetric GM model. The necessary condition for a first-order EWPT is established, and based on this, we numerically investigate the parameter space that can support such a phase transition. Moreover, we consider the GW signals produced by a strong first-order EWPT. Notably, most of the parameter points that yield a strong GW signal exhibit large scalar quartic couplings and significant trilinear couplings between the Higgs boson and DM candidates, which also result in large DM-nucleon scattering cross-sections that are within the reach of current and future direct detection experiments. However, GW measurement can help to explore the Higgs resonant region with $m_H\sim m_h/2$ or $m_5\sim m_h/2$ where the direct detection also loss its sensitivity.

\begin{acknowledgments}
S. Xu and Y. Wu are supported by the National Natural Science Foundation of China (NNSFC) under grant No.~12305112. C.-T. Lu are supported by the NNSFC under grant No.~12335005 and the Special funds for postdoctoral overseas recruitment, Ministry of Education of China. The authors gratefully acknowledge the valuable discussions and insights provided by the members of the China Collaboration of Precision Testing and New Physics (CPTNP).
\end{acknowledgments}

\appendix
\section{The Feynman Rules for the Scalar Interactions in $Z_2$ symmetric GM model}
\label{app:fr}

\subsection{Interactions involving the SM Higgs $h$}

\begin{align}
\frac{g_{HHh}}{v_\phi} &= g_{HHhh} = (4\lambda_2 - 2\lambda_5), \\
\frac{g_{H_3^0H_3^0h}}{v_\phi} &= \frac{g_{H_3^+H_3^-h}}{v_\phi} = g_{H_3^0H_3^0hh} = g_{H_3^+H_3^-hh} = (4\lambda_2 - \lambda_5), \\
\frac{g_{H_5^0H_5^0h}}{v_\phi} &= \frac{g_{H_5^+H_5^-h}}{v_\phi} = \frac{g_{H_5^{++}H_5^{--}h}}{v_\phi}= g_{H_5^0H_5^0hh} = g_{H_5^+H_5^-hh} =g_{H_5^{++}H_5^{--}hh}= (4\lambda_2 + \lambda_5).
\end{align}

\subsection{Interactions among $Z_2$-odd scalars}
\begin{align}
g_{HHHH} &= 3g_{HHH_3^0H_3^0}=3g_{HHH_3^+H_3^-}=8(\lambda_3+3\lambda_4), \\
g_{HHH_5^0H_5^0} &= g_{HHH_5^+H_5^-}=g_{HHH_5^{++}H_5^{--}}=8(\lambda_3+\lambda_4), \\
g_{HH_3^0H_3^0H_5^0} &= -2g_{HH_3^+H_3^-H_5^0} = -\frac{4\sqrt{2}}{3}\lambda_3, \\
g_{HH_3^0H_3^\mp H_5^\pm} &= \pm 2i\sqrt{\frac{2}{3}}\lambda_3, \\
g_{HH_3^\pm H_3^\pm H_5^{\mp\mp}} &= -\frac{4}{\sqrt{3}}\lambda_3, \\
g_{HH_5^0H_5^0H_5^0} &= 2g_{HH_5^0H_5^+H_5^-} = -g_{HH_5^0H_5^{++}H_5^{--}} = 4\sqrt{2}\lambda_3, \\
g_{HH_5^\pm H_5^\pm H_5^{\mp\mp}} &= -4\sqrt{3}\lambda_3, \\
g_{H_3^0H_3^0H_3^0H_3^0} &= 3g_{H_3^0H_3^0H_3^+H_3^-} = 12(\lambda_3 + 2\lambda_4), \\
g_{H_3^0H_3^0H_5^0H_5^0} &= \frac{4}{3}(\lambda_3+6\lambda_4), \\
g_{H_3^0H_3^0H_5^+H_5^-} &= 4(\lambda_3+2\lambda_4), \\
g_{H_3^0H_3^0H_5^{++}H_5^{--}} &= 4(3\lambda_3+2\lambda_4), \\
g_{H_3^0H_5^0H_3^\pm H_5^\mp} &= \mp i\frac{4}{\sqrt{3}}\lambda_3, \\
g_{H_3^0H_3^\pm H_5^\pm H_5^{\mp\mp}} &= \pm 4\sqrt{2}i\lambda_3, \\
g_{H_3^+H_3^+H_3^-H_3^-} &= 8(\lambda_3 + 2\lambda_4), \\
g_{H_3^\pm H_3^\pm H_5^0 H_5^{\mp\mp}} &= 8\sqrt{\frac{2}{3}}\lambda_3, \\
g_{H_3^\pm H_3^\pm H_5^\mp H_5^\mp} &= 8\lambda_3, \\
g_{H_3^+H_3^-H_5^0H_5^0} &= \frac{4}{3}(7\lambda_3+6\lambda_4), \\
g_{H_3^+H_3^-H_5^+H_5^-} &= 8(\lambda_3 + \lambda_4), \\
g_{H_3^+H_3^-H_5^{++}H_5^{--}} &= 4(\lambda_3 + 2\lambda_4), \\
g_{H_5^0H_5^0H_5^0H_5^0} &= 3g_{H_5^0H_5^0H_5^+H_5^-} = 3g_{H_5^0H_5^0H_5^{++}H_5^{--}} = 3g_{H_5^+H_5^-H_5^{++}H_5^--}= 12(\lambda_3 + 2\lambda_4), \\
g_{H_5^+H_5^+H_5^-H_5^-} &= g_{H_5^{++}H_5^{++}H_5^{--}H_5^{--}} = 8(\lambda_3 + 2\lambda_4).
\end{align}

\bibliographystyle{bibsty}
\bibliography{references}

\end{document}